\documentclass[aps,pra,twocolumn,floatfix,groupedaddress,superscriptaddress,footinbib,notitlepage,showpacs]{revtex4-1}
\usepackage{graphicx,graphics,times,bm,bbm,bbold,amssymb,amsmath,amsfonts,dsfont,hyperref,color}
\hypersetup{colorlinks,linkcolor=blue,citecolor=blue,urlcolor=blue}
\newcommand{\lv}{\left\Vert}
\newcommand{\rv}{\right\Vert}
\newcommand{\ignore}[1]{}
\newcommand{\mosh}[1]{\partial_{x}#1}

\newcommand{\ket}[1]{|#1\rangle}
\newcommand{\bra}[1]{\langle #1|}

\newcommand{\ry}{ y }

\newtheorem{corollary}{Corollary}
\newtheorem{lemma}{Lemma}
\newtheorem{theorem}{Theorem}
\let\oldsqrt\sqrt
\def\sqrt{\mathpalette\DHLhksqrt}
\def\DHLhksqrt#1#2{%
\setbox0=\hbox{$#1\oldsqrt{#2\,}$}\dimen0=\ht0
\advance\dimen0-0.2\ht0
\setbox2=\hbox{\vrule height\ht0 depth -\dimen0}%
{\box0\lower0.4pt\box2}}

\usepackage[T1]{fontenc} 
\usepackage{newtxtext,newtxmath}

\DeclareFontFamily{OT1}{pzc}{}
\DeclareFontShape{OT1}{pzc}{m}{it}%
              {<-> s * [1.25] pzcmi7t}{}
\DeclareMathAlphabet{\mathpzc}{OT1}{pzc}%
                                 {m}{it}

\begin{document}

\title{Continuity of the quantum Fisher information}

\author{A. T. Rezakhani}
\email{rezakhani@sharif.edu}
\affiliation{Department of Physics, Sharif University of Technology, Tehran 14588, Iran}
\author{M. Hassani}
\affiliation{Department of Physics, Sharif University of Technology, Tehran 14588, Iran}
\author{S. Alipour}
\affiliation{QTF Center of Excellence, Department of Applied Physics, Aalto University, FI-00076 Aalto, Finland}

\begin{abstract}
In estimating an unknown parameter of a quantum state the quantum Fisher information (QFI) is a pivotal quantity, which depends on the state and its derivate with respect to the unknown parameter. We prove the continuity property for the QFI in the sense that two close states with close first derivatives have close QFIs. This property is completely general and irrespective of dynamics or how states acquire their parameter dependence and also the form of parameter dependence---indeed this continuity is basically a feature of the classical Fisher information that in the case of the QFI naturally carries over from the manifold of probability distributions onto the manifold of density matrices. We demonstrate that in the special case where the dependence of the states on the unknown parameter comes from one dynamical map (quantum channel), the continuity holds in its reduced form with respect to the initial states. In addition, we show that when one initial state evolves through two different quantum channels, the continuity relation applies in its general form. A situation in which such scenario can occur is an open-system metrology where one of the maps represents the ideal dynamics whereas the other map represents the real (noisy) dynamics. In the making of our main result, we also introduce a regularized representation for the symmetric logarithmic derivative which works for general states even with incomplete rank, and it features continuity similarly to the QFI.
\end{abstract}
\maketitle

\section{Introduction}
\label{sec:intro}

Estimation of unknown parameters of a system is an essential task for almost all branches of science and technology. Evidently almost any estimation would entail errors due to various factors such as imperfection of measurement devices or natural stochasticity of the event in question. As a result, estimated values are usually inaccurate. It is of fundamental and practical importance to see what optimal accuracy laws of physics allow in principle. This question of fundamental attainable accuracy in metrology can be addressed by the Cr\'amer-Rao bound \cite{Braunstein-Caves:QFI},
\begin{equation}
\delta  x \geqslant\big(M\mathpzc{F}^{(\mathrm{C})}_{x}\big)^{-1/2},
\label{eq:crb-c}
\end{equation}
where ``$x$'' represents the unknown parameter of interest in a system, $\delta  x $ is the estimation error (i.e., the standard deviation of an unbiased estimator), $M$ is the number of independent repetitions of the estimation protocol, and measurements are performed on an $N$-particle `probe' system. Here the key concept is the (classical) Fisher information, $\mathpzc{F}^{(\mathrm{C})}_{x}(\{p\})$, defined as
\begin{equation}
\mathpzc{F}^{(\mathrm{C})}_{x}\big(\{p\}\big)=\int_{\mathpzc{D}_{\ry}} \mathrm{d}\ry \big(\mosh p(\ry | x )\big)^2/{p(\ry | x )},
\label{CFI-text}
\end{equation}
where $p(\ry| x )$ is the conditional probability for obtaining the value $\ry$ given that the exact value of the parameter is $x$, and $\mathpzc{D}_{\ry}$ is the domain of admissible $\ry$'s. One can see that $\mathpzc{F}_{x}^{(\mathrm{C})}$ scales as $O(N)$ under the assumption that the joint probability for the outcomes of measurements on the $N$-particle probe system is factorized, i.e., the $N$ outcomes are independent and identically-distributed (i.i.d.) random variables. This scaling is called the shot-noise limit \cite{Braunstein-Caves:QFI}.

In quantum metrology, a measurement scenario is described by a set of positive operators $ \lbrace\Pi _{\ry}\rbrace $ which have the completeness property $\int_{\mathpzc{D}_{\ry}} \mathrm{d}\ry\,\Pi_{\ry}=\mathbbmss{I}$, where $\mathbbmss{I}$ is the identity operator. If $\varrho(x)$ denotes the state of the system to be measured, then the probability $p(\ry| x )$ is given by the Born rule  $p(y|x)=\mathrm{Tr}[\varrho(x)\,\Pi_{\ry}]$. Here an important quantity is the symmetric logarithmic derivative (SLD), which is a Hermitian operator $L_{\varrho}$ defined through the (Lyapunov) equation \cite{book:Hayashi}
\begin{equation}
\mosh \varrho=(L_{\varrho}\varrho+\varrho L_{\varrho})/2,
\label{eq:SLD}
\end{equation}
where we have adopted the shorthand $\varrho$ for $\varrho(x)$. The SLD has the following integral representation for \textit{full-rank} density matrices \cite{Paris:tut,Wang}:
\begin{equation}
L_{\varrho} =2\int_{0}^{\infty}\mathrm{d}s~ e^{-s\varrho}~\mosh\varrho~e^{-s\varrho}.
\label{eq:SLD-int}
\end{equation}
Optimizing the Fisher information---attributed to the probabilities obtained through measurements in a quantum metrology scenario---over all measurements yields the quantum Fisher information (QFI)
\begin{equation}
\mathpzc{F}^{(\mathrm{Q})}_{x}(\varrho)=\mathrm{Tr}[\varrho\, L_{\varrho}^2]\overset{(\ref{eq:SLD})}{=}\mathrm{Tr}[\mosh\varrho\, L_{\varrho}] .
\label{def:qfi}
\end{equation}
Thus the quantum Cr\'amer-Rao bound
\begin{equation}
\delta  x \geqslant\big(M\mathpzc{F}^{(\mathrm{Q})}\big)^{-1/2}
\end{equation}
gives the achievable minimum estimation error \cite{Braunstein-Caves:QFI,Paris:tut}, where we have used the lighter notation $\mathpzc{F}^{(\mathrm{Q})}$ for the QFI associated with $x$---we shall adopt this notation throughout the paper. Since a major focus of
quantum metrology is to find good states and measurements which can yield largest quantum or classical Fisher information, we consider $M=1$ for simplicity.

Similarly to the classical case, in Eq. (\ref{def:qfi}), the state $\varrho$ denotes the state of the probe system, usually comprised of $N$ systems (each of which with the Hilbert space $\mathpzc{H}$, hence $\varrho\in\mathpzc{S}(\mathpzc{H}^{\otimes N})$, where $\mathpzc{S}$ is the linear space of linear operators on $\mathpzc{H}^{\otimes N}$), on which a measurement strategy is performed. Bearing this in mind,  however, it can be expected that existence of quantum features may enhance metrology. Particularly, it has been shown that a quantum mechanical enhancement in the form of $\mathpzc{F}^{(\mathrm{Q})}=O(N^2)$ scaling (the Heisenberg limit) can be achieved by employing manybody quantum correlations (in particular, entanglement) \cite{Lloyd-GM,Maccone:PRL}, manybody interactions \cite{Boixo-etal:PRL07}, or nonlinearities \cite{qcorrelation}. In fact, a considerable part of the existing literature on quantum metrology concerns the scaling of the QFI with the probe size, under various conditions, in closed- and open-system metrology scenarios \cite{Escher:Nature,Guta:NC,us,us-benatti}. For a review of this subject, see, e.g., Refs. \cite{Lloyd,Toth}.

Despite this understanding, not much is yet known about specific properties of the QFI \cite{book:Hayashi,book:Petz,Toth}. For example, one can point to \textit{convexity} \cite{Fujiwara,Toth-2,convex-roof}, which recently has been shown to hold for the QFI in the following extended sense \cite{us-convex}:
\begin{equation}
\mathpzc{F}^{(\mathrm{Q})}\big(\textstyle{\sum_a} p_a \varrho_a \big) \leqslant \mathpzc{F}^{(\mathrm{C})}(\{p_a\}) +\textstyle{\sum_a} p_a \mathpzc{F}^{(\mathrm{Q})}(\varrho_a),
\label{ext-conv}
\end{equation}
which reduces to the ordinary convexity property when $p_a$s do not depend on the unknown parameter.

Another essential property to look into is \textit{continuity}. This property (in some sense) has already been shown to hold for, e.g., the von Neumann entropy \cite{cont-S,Audenaert-1,book:Nielsen,Shirokov-1}, quantum conditional entropy \cite{Alicki-Fannes,Winter}, quantum relative entropy and mutual information \cite{Winter,Audenaert-rel,Shirokov-2,cont-asym}, quantum discord \cite{cont-dis}, and some (entropy-based) entanglement measures and quantum channel capacities \cite{Horod,Nielsen-cont,Vidal-cont,Horod-2,Leung-Smith}. For the QFI, however, this property thus far has not been studied in a full generality---for special cases, see Refs. \cite{Kolod,Saf} and remark (vi) of Sec. \ref{sec:main}. This paper is to bridge this gap.

Let us start with an observation about the (classical) Fisher information (\ref{CFI-text}). For two conditional probability distributions $\{p(y|x)\}$ and $\{q(y|x)\}$, we can show that 
\begin{align}
\label{ext-CFI}
&\Big\vert\mathpzc{F}^{(\mathrm{C})}(\{p\})-\mathpzc{F}^{(\mathrm{C})}(\{q\})\Big\vert\leqslant f^{(\mathrm{C})} \int_{\mathpzc{D}_{ y }}\mathrm{d}y\,  \big \vert p(y|x)-q(y|x)\big\vert\nonumber\\ 
&\,+ g^{(\mathrm{C})} \int_{\mathpzc{D}_{ y }}\mathrm{d}y\, \big\vert \mosh p(y|x)-\mosh q(y|x)\big\vert ,
\end{align}
with suitable $f^{(\mathrm{C})}$ and $g^{(\mathrm{C})}$---see appendix \ref{app:CFI-proof} for derivation. Appearance of both the distance of the probability distributions and the distance of their derivatives can be justified by noting that, strictly speaking, the Fisher information depends on the state as well as its derivative. This property should be contrasted with the continuity in its \textit{reduced} sense, discussed in Refs. \cite{Kolod,Saf}, in which under some conditions two `close' states have close Fisher informations. Motivated by this observation, here we derive a continuity relation for the QFI (and similarly for the SLD). This continuity is general in that it is independent of the underlying dynamics for the probe system or how the dependence on the unknown parameter is acquired by the probe state (hence, our relation can be considered as a \textit{kinematical} relation). 

\begin{figure}[tp]
\includegraphics[scale=.25]{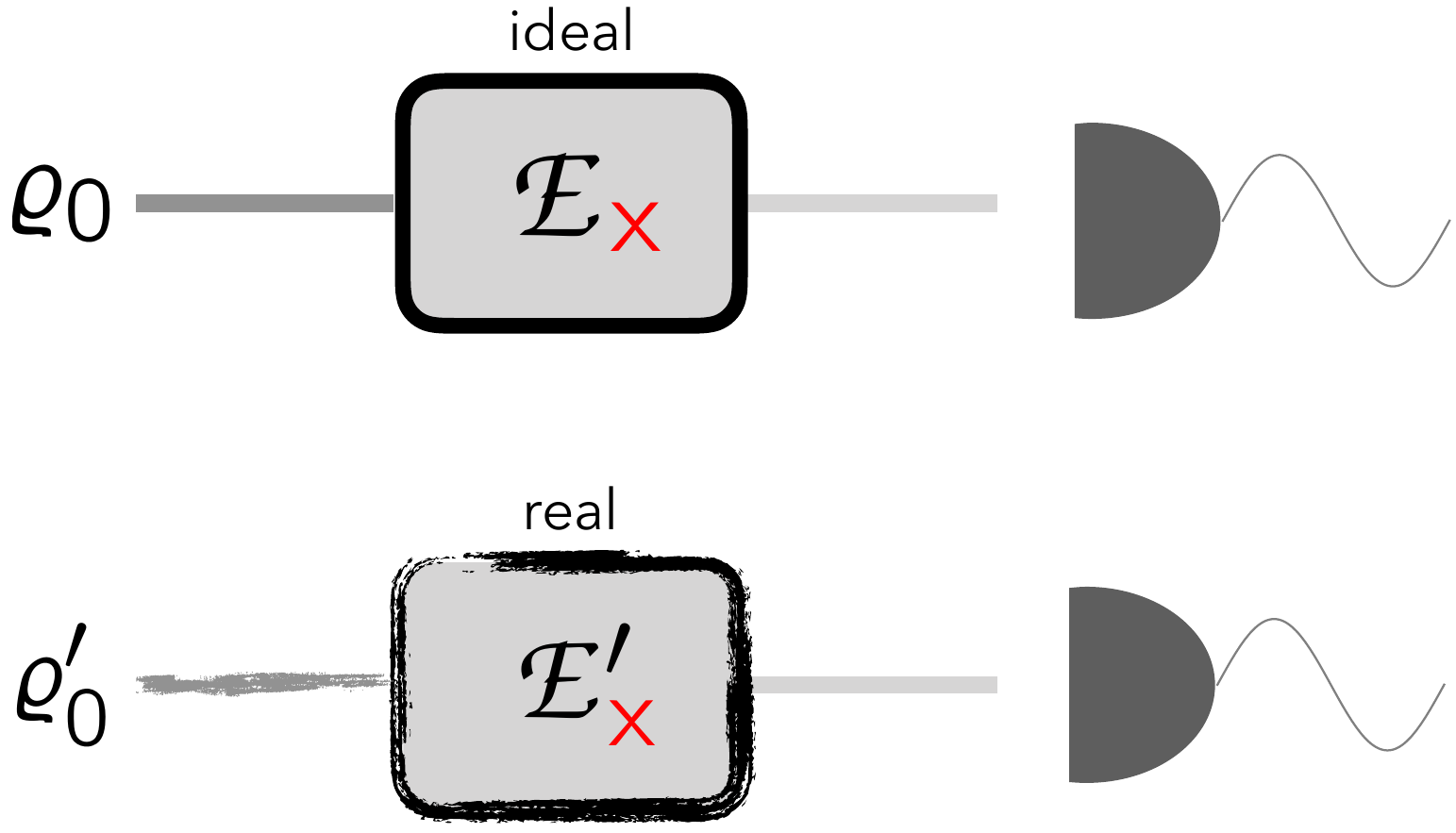}
\caption{Schematic of a general estimation scenario for \textit{dynamics}. (Top): ideal case, where the prepared state $\varrho_{0}$ and the parameter-dependent map $\mathpzc{E}_{\, x }$ (as well as the measurement operation) are assumed ideal (noiseless). (Bottom): real case, where some noise has affected the scenario and changed the initial state $\varrho_{0} \to \varrho'_{0}$ and the dynamics $\mathpzc{E}_{x} \to \mathpzc{E}'_{x}$. Note that $\mathpzc{E}_{x}$ and $\mathpzc{E}'_{x}$ can be by construction general \textit{open-system} dynamics, described by completely-positive, trace-preserving quantum maps or channels \cite{book:Nielsen}. We, however, call $\mathpzc{E}_{x}$ ``ideal'' or ``noiseless'' in the sense that this is what we designed originally; and $\mathpzc{E}'_{x}$ ``real'' or ``noisy'' in the sense that some extra \textit{uncontrollable} or \textit{unaccounted-for} source of noise has changed the designed $\mathpzc{E}_{x}$ to $\mathpzc{E}_{x}'$.}
\label{fig:fig-1}
\end{figure}

Having a continuity relation (in some sense) for the QFI not only is of fundamental importance per se, but also it enables one to investigate relative robustness of metrological scenarios against various sources of ``noise''---in the specific sense illustrated and explained in Fig. \ref{fig:fig-1}. For example, in building fault-tolerant quantum computation \cite{Martinis} and robust quantum sensing with strongly interacting probes \cite{Nemoto}, a continuity relation may have important implications. In addition, noting that in general computing the QFI for systems of large size is even numerically formidable, the continuity relation may allow us to see whether a given state $\varrho_{0}$ or $\varrho_{0}^{\prime}$) may be useful (or useless) for a metrological scenario, without the need to compute the QFI explicitly. Hence, such a relation can offer a significant reduction in complexity of metrology.

We remark that the QFI plays important roles in several other subjects too. For example, it has been shown that when two general density matrices are compared through the quantum fidelity or the closely-related Bures distance (or the Fubini-Study metric in the case of pure states), the QFI takes the role of an information-theoretic metric when the matrices depend continuously on a parameter \cite{Sommers,Paris:tut}. As a result, it seems natural that the QFI may also have intimate connection with quantum phase transitions \cite{Paris-Zanardi,Campos}. In addition, recently the QFI has been employed in high-energy physics and gravity to study the holography property; see, e.g., Ref. \cite{holography}. Having said these, one can anticipate that the utility of continuity relations may naturally go far beyond quantum metrology.

The structure of this paper is as follows. In Sec. \ref{sec:pre} we provide some preliminary norm relations used throughout the paper. In Sec. \ref{sec:main} we lay out our main result and prove it (relegating parts of the proof to several appendices). Section \ref{sec:examples} illustrates our results with three examples or special cases. We conclude and summarize in Sec. \ref{sec:summary}. 

\section{Necessary relations}
\label{sec:pre}

In this section, we establish the preliminary definitions and necessary relations required for proving our main result. 

Let us begin by reminding the definitions of the $p$-norm ($p\in[1,\infty)$) of linear operators. For a linear operator $A$ acting on a linear (finite-dimensional) space, it is defined that $\Vert A\Vert_{p}=(\mathrm{Tr}[|A|^p])^{1/p}$, where $|A|=\sqrt{A^{\dag}A}$ \cite{book:Bhatia,book:Watrous}. \ignore{For brevity, we omit subscript $\infty$ from the $\infty$-norm (the standard sup-operator norm) and simply write $\Vert \,\Vert$ instead.}The $p$-norms have numerous appealing properties. One particular property useful for our purpose in this paper is the following duality between the $1$-norm (or trace norm) and the $\infty$-norm (or standard norm) \cite{book:Watrous,Lidar,Hindi,Wolf}:
\begin{equation}
\Vert A\Vert_1 =\sup_{B\neq0}\frac{|\mathrm{Tr}[B^{\dag}A]|}{\Vert B\Vert _{\infty} },
\end{equation}
from whence we obtain this useful inequality
\begin{equation}
|\mathrm{Tr}[BA]|\leqslant \Vert A\Vert_{1} \,\Vert B^{\dag}\Vert _{\infty} \leqslant \Vert A\Vert_{1} \,\Vert B^{\dag}\Vert_{1}.
\label{duality}
\end{equation}
Note that $\Vert A\Vert_{\infty}=\lim_{p\to\infty}\Vert A\Vert_{p}=\sup_{i}s_{i}(A)$, where $\{s_{i}(A)\}$ are the eigenvalues of $|A|$ (singular values of $A$). The last inequality above is a special case of the property
\begin{equation}
\Vert A\Vert_{q}\leqslant\Vert A\Vert_{p},~~1\leqslant p\leqslant q\leqslant\infty.
\label{eq:p-ni}
\end{equation}
Another useful property is the submultiplicativity in the form
\begin{equation}
\Vert A B\Vert_{p}\leqslant \Vert A\Vert_{p}\, \Vert B\Vert_{p},
\end{equation}
for any $p$. In addition, we also have  
\begin{align}
\Vert AB\Vert_{1} &\leqslant \Vert A\Vert _{\infty} \,\Vert B\Vert_{1},\Vert A\Vert_{1}\,\Vert B\Vert _{\infty}, 
\label{norm-1-1-inf}\\
\Vert ABC\Vert_{p} &\leqslant \Vert A\Vert _{\infty} \,\Vert B\Vert_{p} \,\Vert C\Vert _{\infty}.
\label{new-ineq}
\end{align}

Let $A$, $B$, $A'$, and $B'$ be linear operators. We have
\begin{align}
\Vert AB-A' B'\Vert_{p} &= \Vert (AB-A'B) + (A'B-A'B')\Vert_{p} \nonumber\\
& \leqslant \Vert (A-A')B\Vert_{p} + \Vert A'(B-B')\Vert_{p},
\label{imp-ineq}
\end{align}
which is akin to the inequality for classical probability distributions---see Eq. (\ref{abs-diff}). This, noting Eq. (\ref{norm-1-1-inf}), yields
\begin{align}
\Vert AB-A' B'\Vert_{1} \leqslant& \Vert A-A'\Vert _{\infty}\, \Vert B\Vert_{1} + \Vert A'\Vert_{1}\, \Vert B-B'\Vert _{\infty},
\label{inequality-1} \\
\Vert AB-A' B'\Vert_{1} \leqslant& \Vert A-A'\Vert_{1}\,\Vert B\Vert _{\infty}  + \Vert A'\Vert _{\infty}\,  \Vert B-B'\Vert_{1}.
\label{inequality-2} 
\end{align}
As a result, when we take $A=K_{1}X$, $B=K_{1}^{\dag}$, $A'=K_{2}X$, and $B'=K_{2}^{\dag}$, we obtain
\begin{align}
\Vert K_{1}X K_{1}^{\dag} - K_{2}X K_{2}^{\dag}\Vert_{\infty} \leqslant&  \Vert K_{1} - K_{2} \Vert _{\infty} \,\Vert X\Vert_{1} \big( \Vert K_{1}\Vert  _{\infty}+ \Vert K_{2}\Vert _{\infty} \big),
\label{ineq-1}
\end{align}
where we have also used $\Vert O^{\dag}\Vert_{p}=\lv O\rv_{p}$ (valid for any $p$-norm on linear spaces). For the more general case of $\Vert K_{1}X_{1} K_{1}^{\dag} - K_{2}X_{2} K_{2}^{\dag}\Vert _{\infty}$, from Eq. (\ref{ineq-1}) we have 
\begin{align}
&\Vert K_{1}X_{1} K_{1}^{\dag} - K_{2}X_{2} K_{2}^{\dag}\Vert _{\infty}
\leqslant \Vert X_{1} -X_{2} \Vert_{1}\, \Vert K_{1}\Vert^{2}_{\infty} \nonumber\\
&+\Vert K_{1} - K_{2} \Vert _{\infty} \,\Vert X_{2}\Vert_{1}\big( \Vert K_{1}\Vert _{\infty} + \Vert K_{2}\Vert _{\infty} \big).
\label{important-ineq}
\end{align}
Another identity which will be important for our analysis is as follows:
\begin{align}
e^{A}-e^{B}=\int_{0}^{1}\mathrm{d}\tau~ e^{\tau A}(A-B)e^{(1-\tau)B},
\label{ineq4}
\end{align}
for any pair of linear operators defined on a given linear space. To prove this, we choose $V(\tau)=e^{\tau A}e^{-\tau B}$ and use $V(\mathbbmss{I})-\mathbbmss{I}=\int_{0}^{1}\mathrm{d}\tau~ \mathrm{d}V(\tau)/\mathrm{d}\tau$. An immediate consequence is
\begin{equation}
\Vert e^{A} -e^{B} \Vert_{p} \leqslant \Vert A-B\Vert_{p} \int_{0}^{1}\mathrm{d}\tau~\Vert e^{\tau A}\Vert_{p}\, \Vert e^{(1-\tau)B}\Vert_{p}.
\label{aa}
\end{equation}
A useful special case is when $A=-s\varrho$, $B=-s\sigma$, $s\geqslant 0$, in which $\varrho$ and $\sigma$ are two quantum states (density matrices) of a given system. We first note that
\begin{equation}
\Vert e^{-s\varrho}\Vert_{\infty} =e^{-s\lambda_{\min}(\varrho)},
\label{norm-e-rho}
\end{equation}
where $\lambda_{\min}(\varrho)$ is the smallest eigenvalue of $\varrho$. Using this relation, we can calculate the integral in Eq. (\ref{aa}), from whence
\begin{equation}
\Vert e^{-s\varrho}-e^{-s\sigma}\Vert_{\infty}  \leqslant \frac{e^{-s\lambda_{\min}(\sigma)}- e^{-s\lambda_{\min}(\varrho)}}{\lambda_{\min}(\varrho)- \lambda_{\min}(\sigma)}\Vert \varrho-\sigma\Vert_{1}.
\label{ineq:esr}
\end{equation}
Applying a similar method to Eq. (\ref{eq:SLD-int}) yields
\begin{equation}
\Vert L_{\varrho}\Vert_{\infty} \leqslant  \Vert \mosh\varrho\Vert_{\infty} /\lambda_{\min}(\varrho).
\label{norm-L}
\end{equation}

Since in our derivation later in the paper it is necessary to upperbound $\Vert L_{\varrho}\Vert_{\infty}$, this relation will be useful. However, when $\varrho$ is incomplete-rank, this upper bound become vacuous ($\infty$). In fact, as we carefully argue in appendix \ref{app:SLD}, in some particular situations concerning the incomplete-rank case this divergence shows up due to the integral representation (\ref{eq:SLD-int}). To remedy this issue, here we utilize an inherent freedom of the SLD (\ref{eq:SLD}) in the QFI (see also appendix \ref{app:SLD}) in order to introduce a \textit{regularized} representation $\mathpzc{L}_{\varrho}$ (called ``r-SLD'') as follows:
\begin{align}
\mathpzc{L}_{\varrho}=&2\int_{0}^{\infty}\mathrm{d}s~e^{-s \widetilde{\varrho}}~\mosh\varrho ~e^{-s \widetilde{\varrho}} +2\big(P_{\varrho}~\mosh P_{\varrho}~P_{\varrho}^{\perp} +P_{\varrho}^{\perp}~\mosh P_{\varrho}~P_{\varrho}\big),
\label{sing-L-2}
\end{align}
where $\widetilde{\varrho}$ is the (invertible) \textit{restriction} of $\varrho$ onto $P_{\varrho}\mathpzc{H}^{\otimes N}P_{\varrho}$, and $P_{\varrho}$ ($P_{\varrho}^{\perp}$) is the projector onto the support (null subspace) of the density matrix $\varrho$---noting $ P_{\varrho}+P_{\varrho}^{\perp}=\mathbbmss{I}$. In fact, although $P^{\perp}_{\varrho} L_{\varrho} P^{\perp}_{\varrho}$ is divergent (appendix \ref{app:SLD}), we have $P^{\perp}_{\varrho} \mathpzc{L}_{\varrho} P^{\perp}_{\varrho}=0$ (because $P^{\perp}_{\varrho}e^{-s\widetilde{\varrho}} = e^{-s\widetilde{\varrho}}P^{\perp}_{\varrho} =0$, whereas $P^{\perp}_{\varrho}e^{-s\varrho} = e^{-s\varrho} P^{\perp}_{\varrho} = P^{\perp}_{\varrho}$). Note that although $\mathpzc{L}_{\varrho}$ differs in the form with $L_{\varrho}$, it is straightforward to see that (see appendix \ref{app:SLD}) they both satisfy Eq. (\ref{eq:SLD}), and as long as the QFI is concerned, these two quantities are equivalent, 
\begin{equation}
\mathpzc{F}^{(\mathrm{Q})}_{x}(\varrho)=\mathrm{Tr}[\varrho\, L_{\varrho}^2] = \mathrm{Tr}[\varrho\, \mathpzc{L}_{\varrho}^2].
\label{def:qfi-2}
\end{equation}

Note that in Eq. (\ref{sing-L-2}), $\partial_{x}P_{\varrho}$ becomes undefined exactly at $x$ values where the rank of $\varrho$ changes. Thus $\mathpzc{L}_{\varrho}$ is well-defined everywhere expect at rank-changing points. Excluding such problematic points, from Eq. (\ref{sing-L-2}) we obtain the following upper bound on the norm of the r-SLD operator:    
\begin{align}
\Vert \mathpzc{L}_{\varrho}\Vert_{\infty} \leqslant \Vert \mosh\varrho\Vert_{\infty} / \lambda_{\min}(\widetilde{\varrho}) +4\Vert\mosh P_{\varrho}\Vert_{\infty},
\label{sing-L-norm} 
\end{align}
where we have used the fact that the standard norm of projectors is unity ($\Vert P_{\varrho}\Vert_{\infty}=\Vert P_{\varrho}^{\perp}\Vert_{\infty} =1$). Obviously, when $\varrho$ is full-rank, this bound reduces to the bound (\ref{norm-L}).

\section{Continuity relation}
\label{sec:main}

We start with a remark regarding the notations ``$L_{\varrho}$'' and ``$\mathpzc{F}^{(\mathrm{Q})}(\varrho)$.'' It should be understood that putting ``$\varrho$'' here is for brevity and does not necessarily imply that the SLD or QFI are \textit{functions} (in the conventional mathematical sense) of $\varrho$ alone. In fact, from the definitions (\ref{eq:SLD-int}) and (\ref{def:qfi}) it is evident that both SLD and QFI are functions (strictly speaking, functionals) of $\varrho$ and $\partial_{x}\varrho$. Thus it is more appropriate to define the shorthand boldface symbol $\bm{\varrho}=(\varrho,\partial_{x}\varrho)$ and hereafter represent the SLD, r-SLD, and QFI with $L_{\bm{\varrho}}$, $\mathpzc{L}_{\bm{\varrho}}$, and $\mathpzc{F}^{(\mathrm{Q})}(\bm{\varrho})$, respectively. Indeed mathematically speaking, the QFI is a function defined on the \textit{tangent bundle} \cite{Fujiwara,book:nakahara} of the manifold of the density matrices. The following theorem encompasses our main result.

\begin{widetext}
\begin{theorem}
\label{thm:1}
For any pair of density matrices $\varrho$ and $\sigma$ (depending on the same unknown parameter $x$, although irrespectively of their specific dependence on this parameter) defined on the Hilbert space $\mathpzc{H}^{\otimes N}$, we have 
\begin{align}
\big|\mathpzc{F}^{(\mathrm{Q})}(\bm{\varrho})-\mathpzc{F}^{(\mathrm{Q})}(\bm{\sigma})\big| \leqslant  f^{(\mathrm{Q})} \Vert \varrho-\sigma\Vert_{1} + g^{(\mathrm{Q})} \Vert \mosh\varrho - \mosh\sigma \Vert_{1},
\label{continuity-2}
\end{align}
where
\begin{align}
f^{(\mathrm{Q})} = & \frac{\Vert \mosh\sigma\Vert_{1}}{\lambda_{\min}(\widetilde{\varrho})\,\lambda_{\min}(\widetilde{\sigma})}
\left(\Vert \mosh\sigma\Vert_{1}+32(\Vert \mosh P_{\varrho}\Vert_{\infty} +\Vert \mosh P_{\sigma}\Vert_{\infty} )+64 \Vert \mosh\sigma\Vert_{1}\frac{\lambda_{\min}(\widetilde{\varrho})+\lambda_{\min}(\widetilde{\sigma})}{\lambda_{\min}(\widetilde{\varrho}) \,\lambda_{\min}(\widetilde{\sigma})} \right),\label{def:f}\\
g^{(\mathrm{Q})} = &   \frac{1}{\lambda_{\min}(\widetilde{\varrho})}\big(\Vert\mosh\varrho\Vert_{1}+\Vert\mosh\sigma\Vert_{1}\big)+4\Vert \mosh P_{\varrho}\Vert_{\infty} +\frac{32\Vert\mosh\sigma\Vert_{1}}{\lambda_{\min}^{2}(\widetilde{\varrho})}. \label{def:g}
\end{align}
Noting that the QFI is a function of both the state and its derivate, Eq. (\ref{continuity-2}) can be interpreted as a ``continuity relation.''
\end{theorem}

\textit{Proof.} From the definition of the QFI (\ref{def:qfi}) [or Eq. (\ref{def:qfi-2})] for two parameter-dependent states $\varrho$ and $\sigma$, we have
\begin{align}
\big|\mathpzc{F}^{(\mathrm{Q})}(\bm{\varrho})-\mathpzc{F}^{(\mathrm{Q})}(\bm{\sigma})\big|
&=\big|\mathrm{Tr}[\mosh\varrho\, \mathpzc{L}_{\bm{\varrho}}-\mosh\sigma\, \mathpzc{L}_{\bm{\sigma}}]\big|\nonumber\\
&\overset{\mathrm{(\ref{duality})}}{\leqslant} \lv \mosh\varrho\, \mathpzc{L}_{\bm{\varrho}}-\mosh\sigma\, \mathpzc{L}_{\bm{\sigma}}\rv_{1}.
\end{align}
If we replace $A=\mosh\varrho$, $B=\mathpzc{L}_{\bm{\varrho}}$, $A^{\prime}=\mosh\sigma$, and $B^{\prime}=\mathpzc{L}_{\bm{\sigma}}$ in Eq. \eqref{inequality-1}, we obtain
\begin{align}
\big|\mathpzc{F}^{(\mathrm{Q})}(\bm{\varrho})-\mathpzc{F}^{(\mathrm{Q})}(\bm{\sigma})\big|\leqslant& \Vert \mosh\varrho -\mosh\sigma \Vert_{1}\, \Vert \mathpzc{L}_{\bm{\varrho}}\Vert_{\infty} +\Vert\mosh\sigma\Vert_{1}\, \Vert  \mathpzc{L}_{\bm{\varrho}}-\mathpzc{L}_{\bm{\sigma}}\Vert_{\infty},
\label{continuity}
\end{align}
Equation (\ref{continuity}) indicates that we still need to calculate $\lv \mathpzc{L}_{\bm{\varrho}}-\mathpzc{L}_{\bm{\sigma}}\rv_{\infty} $ in terms of more primitive quantities (e.g., relevant properties of $\varrho$, $\sigma$, and perhaps their derivatives). We start from the integral representation (\ref{sing-L-2}), whence
\begin{align}
\lv \mathpzc{L}_{\bm{\varrho}}-\mathpzc{L}_{\bm{\sigma}}\rv_{\infty} \leqslant &2\int_{0}^{\infty}\mathrm{d}s\,\big\Vert e^{-s\widetilde{\varrho}}\mosh\varrho~e^{-s\widetilde{\varrho}}-e^{-s\widetilde{\sigma}}\mosh\sigma~ e^{-s\widetilde{\sigma}}\big\Vert_{\infty} + 2 \big\Vert P_{\varrho}~\mosh P_{\varrho}~P_{\varrho}^{\perp}+P_{\varrho}^{\perp}~\mosh P_{\varrho}~P_{\varrho} - P_{\sigma}~\mosh P_{\sigma}~P_{\sigma}^{\perp}-P_{\sigma}^{\perp}~\mosh P_{\sigma}~P_{\sigma}\big\Vert_{\infty}.
\label{ll}
\end{align}
Now if we employ Eq. (\ref{important-ineq}) with $K_1=e^{-s\widetilde{\varrho}}$, $K_2=e^{-s\widetilde{\sigma}}$, $X_1=\mosh\varrho$, and $X_2=\mosh\sigma$, the first term on the right-hand side (RHS) of Eq. (\ref{ll}) becomes upperbounded by 
\begin{align}
 2 \int_{0}^{\infty}\mathrm{d}s~\Big[\Vert \mosh\varrho - \mosh\sigma\Vert_{1}\,\Vert e^{-s\widetilde{\varrho}}\Vert^{2}_{\infty}  +\Vert e^{-s\widetilde{\varrho}}- e^{-s\widetilde{\sigma}} \Vert_{\infty} \,\Vert \mosh\sigma\Vert_{1}\big( \Vert e^{-s\widetilde{\varrho}}\Vert_{\infty} + \Vert e^{-s\widetilde{\sigma}}\Vert_{\infty} \big)\Big].
\label{l-l}
\end{align}
The above integrals can be upperbounded by using Eqs. (\ref{norm-e-rho}) and (\ref{ineq:esr}), which gives
\begin{align}
 \frac{1}{\lambda_{\min}(\widetilde{\varrho})}\Vert \mosh\varrho - \mosh\sigma\Vert_{1} +\frac{\Vert \mosh\sigma\Vert_{1}}{\lambda_{\min}(\widetilde{\varrho}) \, \lambda_{\min}(\widetilde{\sigma}) }\Vert \varrho- \sigma\Vert_{1} \label{L-L1}.
\end{align}

For the second term on the RHS of Eq. (\ref{ll}) term, after using Eqs. (\ref{inequality-1}) and (\ref{inequality-2}) twice, we obtain
\begin{align}
\Vert P_{\varrho}~\mosh P_{\varrho}~P_{\varrho}^{\perp} + P_{\varrho}^{\perp}~\mosh P_{\varrho}~P_{\varrho} - P_{\sigma}~\mosh P_{\sigma}~P_{\sigma}^{\perp}-P_{\sigma}^{\perp}~\mosh P_{\sigma}~P_{\sigma}\Vert_{\infty} \leqslant&  2\Big(\big(\Vert \partial_{x}P_{\varrho}\Vert_{\infty} + \Vert \partial_{x}P_{\sigma}\Vert_{\infty} \big) \Vert P_{\varrho} - P_{\sigma} \Vert_{\infty} + \Vert \partial_{x}P_{\varrho} - \partial_{x}P_{\sigma} \Vert_{\infty}\Big).
\end{align}
From Lemma \ref{lemma:1} of appendix \ref{app:pp} we have 
\begin{align}
\Vert P_{\varrho}-P_{\sigma} \Vert_{\infty} &\leqslant \frac{8}{\lambda_{\min}(\widetilde{\varrho}) \, \lambda_{\min}(\widetilde{\sigma})}\Vert \varrho -\sigma\Vert_{1}, \label{bound-p-p}\\
\Vert \partial_{x}P_{\varrho} - \partial_{x}P_{\sigma}\Vert_{\infty} & \leqslant 8 \left(\frac{1}{ \lambda^{2}_{\min}(\widetilde{\varrho})} \Vert \partial_{x}\varrho - \partial_{x}\sigma\Vert_{1} +  2\frac{\big(\lambda_{\min}(\widetilde{\varrho}) + \lambda_{\min}(\widetilde{\sigma})\big)\Vert \partial_{x}\sigma\Vert_{1} }{\lambda^{2}_{\min}(\widetilde{\varrho}) \, \lambda^{2}_{\min}(\widetilde{\sigma})} \Vert \varrho -\sigma\Vert_{1}\right). \label{bound-dp-dp}
\end{align} 
Inserting Eqs. (\ref{bound-p-p}) and (\ref{bound-dp-dp}) back into Eq. (\ref{ll}) yields
\begin{equation}
\Vert \mathpzc{L}_{\bm{\varrho}} - \mathpzc{L}_{\bm{\sigma}}\Vert_{\infty} \leqslant a\Vert \varrho -\sigma\Vert_{1} + b\Vert \partial_{x}\varrho - \partial_{x}\sigma\Vert_{1},
\label{ll-cont}
\end{equation}
where
\begin{align}
a&=\frac{1}{\lambda_{\min}(\widetilde{\varrho}) \, \lambda_{\min}(\widetilde{\sigma})}
\left(\Vert \mosh\sigma\Vert_{1}+32\big(\Vert \mosh P_{\varrho}\Vert_{\infty} +\Vert \mosh P_{\sigma}\Vert_{\infty} \big)+64 \Vert \mosh\sigma\Vert_{1}\frac{\lambda_{\min}(\widetilde{\varrho})+\lambda_{\min}(\widetilde{\sigma})}{\lambda_{\min}(\widetilde{\varrho}) \,  \lambda_{\min}(\widetilde{\sigma})} \right),\label{def:a}\\
b&=\frac{1}{\lambda_{\min}(\widetilde{\varrho})}\Big(1+\frac{32\Vert\mosh\sigma\Vert_{1}}{\lambda_{\min}(\widetilde{\varrho})}\Big). \label{def:b}
\end{align}
\ignore{
or if we use the bound (\ref{norm-P-dot})
\begin{align}
a&=\frac{1}{\lambda_{\min}(\widetilde{\varrho})\lambda_{\min}(\widetilde{\sigma})}
\left(\Vert \mosh\sigma\Vert_{1}+2^{8}\Big(\frac{\Vert \mosh \varrho\Vert}{\lambda_{\min}^{2}(\widetilde{\varrho})} + \frac{\Vert \mosh \sigma\Vert}{\lambda_{\min}^{2}(\widetilde{\sigma})} \Big)+ 2^{6} \Vert \mosh\sigma\Vert_{1}\frac{\lambda_{\min}(\widetilde{\varrho})+\lambda_{\min}(\widetilde{\sigma})}{\lambda_{\min}(\widetilde{\varrho}) \lambda_{\min}(\widetilde{\sigma})} \right), \label{def:a-}\\
b&=\frac{1}{\lambda_{\min}(\widetilde{\varrho})}\Big(1+\frac{32\Vert\mosh\sigma\Vert_{1}}{\lambda_{\min}(\widetilde{\varrho})}\Big).\label{def:b-}
\end{align}
 }
This relation establishes a continuity property for the r-SLD. Due to Eq. (\ref{continuity}), the latter continuity of the r-SLD carries over to the QFI too; doing so, we obtain Eq. (\ref{continuity-2}).

\hfill$\square$

\ignore{
or if we use Eq. (\ref{norm-P-dot})
\begin{align}
f^{(\mathrm{Q})} = & \frac{\Vert \mosh\sigma\Vert_{1}}{\lambda_{\min}(\widetilde{\varrho})\lambda_{\min}(\widetilde{\sigma})}
\left(\Vert \mosh\sigma\Vert_{1}+2^{8}\Big( \frac{\Vert \mosh\varrho \Vert}{\lambda^{2}_{\min}(\widetilde{\varrho})}  + \frac{\Vert \mosh\sigma \Vert}{\lambda^{2}_{\min}(\widetilde{\sigma})} \Big) + 2^{6} \Vert \mosh\sigma\Vert_{1}\frac{\lambda_{\min}(\widetilde{\varrho})+\lambda_{\min}(\widetilde{\sigma})}{\lambda_{\min}(\widetilde{\varrho}) \lambda_{\min}(\widetilde{\sigma})} \right),\label{def:f-}\\
g^{(\mathrm{Q})} = &   \frac{1}{\lambda_{\min}(\widetilde{\varrho})} \big(\Vert\mosh\varrho\Vert_{1}+\Vert\mosh\sigma\Vert_{1}\big)+ \frac{2^{5}\Vert\mosh\varrho\Vert_{1}}{\lambda_{\min}^{2}(\widetilde{\varrho})} +\frac{2^{5}\Vert\mosh\sigma\Vert_{1}}{\lambda_{\min}^{2}(\widetilde{\varrho})}. \label{def:g-}
\end{align}
}
\end{widetext}

\begin{corollary}
If $\eta=\max_{}\{f^{(\mathrm{Q})},g^{(\mathrm{Q})}\}$, we can write Eq. (\ref{continuity-2}) as
\begin{equation}
\big|\mathpzc{F}^{(\mathrm{Q})}(\bm{\varrho})-\mathpzc{F}^{(\mathrm{Q})}(\bm{\sigma})\big| \leqslant \eta\, D(\bm{\varrho},\bm{\sigma}),
\label{continuity-3}
\end{equation}
where $D( \bm{\varrho},\bm{\sigma})=\Vert \varrho-\sigma\Vert_{1} + \Vert \partial_{x}\varrho-\partial_{x}\sigma\Vert_{1}$ is a distance measure.
\end{corollary}

\begin{corollary}
\label{corr:2}
For the case of full-rank density matrices, our Eqs. (\ref{def:f}) and (\ref{def:g}) reduce to
\begin{align}
f^{(\mathrm{Q})} = & \frac{\Vert \mosh\sigma\Vert_{1}^{2}}{\lambda_{\min}(\widetilde{\varrho})\, \lambda_{\min}(\widetilde{\sigma})},\\
g^{(\mathrm{Q})} = &   \frac{1}{\lambda_{\min}(\widetilde{\varrho})}\big(\Vert\mosh\varrho\Vert_{1}+\Vert\mosh\sigma\Vert_{1}\big).
\end{align}
These expressions should be contrasted with their classical counterparts $f^{(\mathrm{C})}$ and $g^{(\mathrm{C})}$ in appendix \ref{app:CFI-proof} [Eqs. (\ref{f-class}) and (\ref{g-class})]. 
\end{corollary}

Several remarks are in order here.

(i). As a caveat, note that the bounds and inequalities we have derived in this paper are not necessarily tight. This can be partly alleviated by replacing $\bm{\varrho} \leftrightarrow \bm{\sigma}$ in the bounds and then taking the minimum of the the two sets of expressions for the bound as a tighter and more appealing substitute (because of the $\bm{\varrho} \leftrightarrow \bm{\sigma}$ symmetry).

(ii). If we restrict the density matrices to a domain (a subspace of $\mathpzc{S}(\mathpzc{H}^{\otimes N})$) on which $f^{(\mathrm{Q})}$ and $g^{(\mathrm{Q})}$ can be bounded (that is, $\eta<\infty$), one can consider our bound (\ref{continuity-2}) as a \textit{Lipschitz continuity relation}---for an introduction to the Lipschitz continuity, see Ref. \cite{book:Bhatia}.

(iii). Although we can still simplify the expressions of Eqs. (\ref{def:a}) and (\ref{def:b}) (by using Eq. (\ref{app:dP}) of appendix \ref{app:pp}), the existing forms are more preferable since they are smaller and also better capture the behavior of the bound (\ref{continuity-2}) for the case of full-rank states.

(iv). When an initial state $\varrho_{0}$ evolves by a general dynamical map $\mathpzc{E}_{x}$ (see also subsections \ref{example:noiseless} and \ref{example:noise}), we can obtain a lower bound on $\lambda_{\min}(\mathpzc{E}_{x}[\varrho_{0}])$---appendix \ref{app:eig}. Although $ \lambda_{\min}(\widetilde{\varrho}) $ and $ \lambda_{\min}(\widetilde{\sigma}) $ are strictly nonzero, they both might become infinitesimally small. This (pathological) case, however, does not impact our main relation because the upper bound in Eq. (\ref{continuity-2}) holds irrespectively of how small $\lambda_{\min}$s are (in particular, in the $ \Vert\varrho -\sigma\Vert_{1}\to 0$ and $ \Vert\mosh\varrho -\mosh\sigma\Vert_{1} \to 0$ limit). In such cases, however, our bound may become vacuous. This phenomenon can be interpreted as the failure of the continuity relation in the sense that the bound (\ref{continuity-2}) diverges. It should, however, be cautioned that such divergence of our bound does not necessary imply that the difference of the QFIs must diverge (although the converse is always true). As a side, it may be relevant to note that if one of the eigenvalues of the density matrix approaches zero, the QFI matrix will fail to ``concentrate'' \cite{Guta}. It could be interesting to see if it is possible to conclude conditions for such failure of the concentration from our continuity relation. We, however, leave this investigation as an open problem.

\begin{figure}[bp]
\includegraphics[scale=0.37]{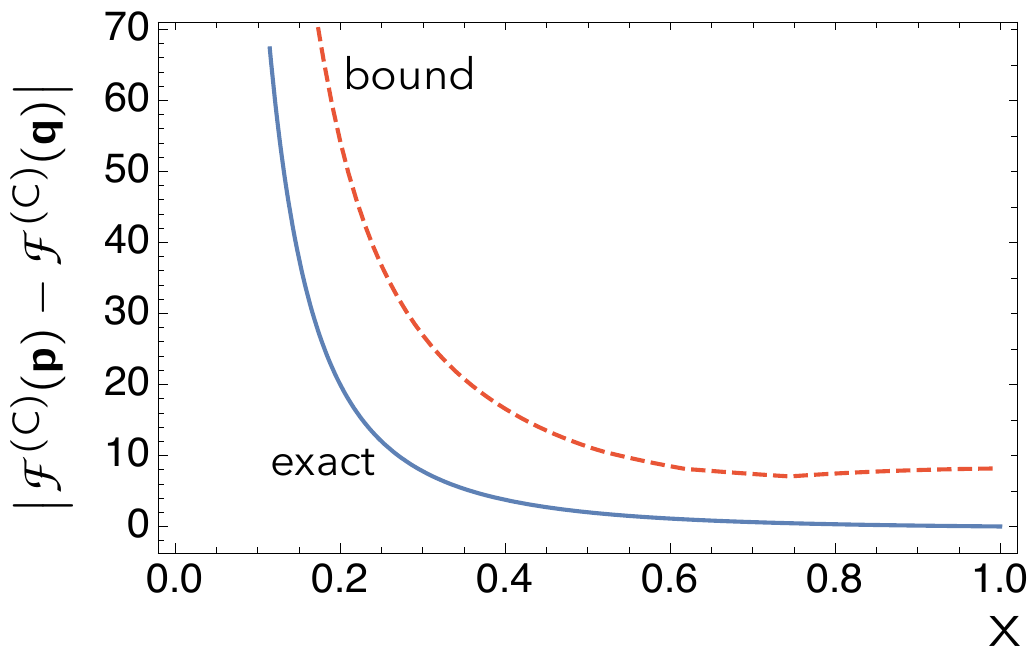}
\caption{$\big|\mathpzc{F}^{(\mathrm{C})}(\mathbf{p})- \mathpzc{F}^{(\mathrm{C})}(\mathbf{q})\big|$ vs. $x$ for the example discussed in remark (v).}
\label{fig:class}
\end{figure}

(v). Singular appearance of $\lambda_{\min}(\widetilde{\varrho})$ and $\lambda_{\min}(\widetilde{\sigma})$ (and their combinations) in $f^{(\mathrm{Q})}$ and $g^{(\mathrm{Q})}$ of the upper bound (\ref{continuity-2}) can be partially justified by comparing these quantities with their classical counterparts $f^{(\mathrm{C})}$ and $g^{(\mathrm{C})}$, where now $p(y|x)$ is replaced with $\varrho$, $\min_{ y \in\mathpzc{D}_{ y }}p(y|x)$ with $\lambda_{\min}(\widetilde{\varrho})$, and $\partial_{x}p(y|x)$ with $\partial_{x}\varrho$. With this recipe, similarities are evident and one may have a better understanding why specific and complex combinations show up in the coefficients $f^{(\mathrm{Q})}$ and $g^{(\mathrm{Q})}$.

As a specific example, consider the problem of estimating a classical parameter $0 \leqslant  x  \leqslant 1$ by some scheme, which has given the measurement results $ y \in\mathpzc{D}_{ y }=[0,2]$. Assume that we estimate the parameter $x$ with two probability distributions $p(y|x)= x \,e^{- x  y }$ and $q(y|x) = e^{- x }  x ^{ y }/ y !$, for $ y \geqslant 0$. We have $\mathpzc{F}^{(\mathrm{C})}(\mathbf{p})=1/ x ^2 $ and $\mathpzc{F}^{(\mathrm{C})}(\mathbf{q})=1/ x $, and thus $|\mathpzc{F}^{(\mathrm{C})}(\mathbf{p}) - \mathpzc{F}^{(\mathrm{C})}(\mathbf{q})|\to \infty$ when $ x \to0$, or equivalently when $p,q\to 0$. However, in this limit we have $|p - q|\to0,|\partial_{x}p - \partial_{x}q|<\infty$ but $f^{(\mathrm{C})},g^{(\mathrm{C})}\to\infty$---see Fig. \ref{fig:class}. This behavior is compatible with the $1/\min p$ and $1/(\min p \,\min q)$ dependence of the coefficients $f^{(\mathrm{C})}$ and $g^{(\mathrm{C})}$ as in appendix \ref{app:CFI-proof} [Eqs. (\ref{f-class}) and (\ref{g-class})].

(vi). After completion of the first version of this work \cite{A-T-R}, we became aware that the special case of closed-system (unitary) metrology has been recently analyzed in the sense of both \textit{reduced} continuity and entanglement in Ref. \cite{Kolod}. A while later, another reference appeared \cite{Saf} wherein ``discontinuities'' of the QFI and Bures metric have been studied in a \textit{different} sense: considering $\mathpzc{F}^{(\mathrm{Q})}_{x}$ as a real-valued function of a single real parameter ($x$), and hence, comparing $\mathpzc{F}^{(\mathrm{Q})}_{x}$ and $\mathpzc{F}^{(\mathrm{Q})}_{ x +\mathrm{d}x}$. There the parametrization of the state is assumed \textit{fixed} in terms of $x$; $\varrho(x)\to\varrho( x +\mathrm{d}x)$. In this reference, it has been argued that when during the change of the estimation parameter there is a rank change for $\varrho(x)$, the QFI is ``discontinuous'' in this particular sense. 

Note that all of these results are compatible with our main message. However, within a more general context, our continuity relation improves upon these few relevant studies in various aspects. Specifically, our bounds are completely general, independent of the way the unknown parameter has entered the description of the state, and apply for any kind of parameter dependence in the state (modulo differentiability).

\section{Examples and special cases} 
\label{sec:examples}

\subsection{Qubit} 
\label{Qubit}

\begin{figure}[tp]
\includegraphics[scale=0.29]{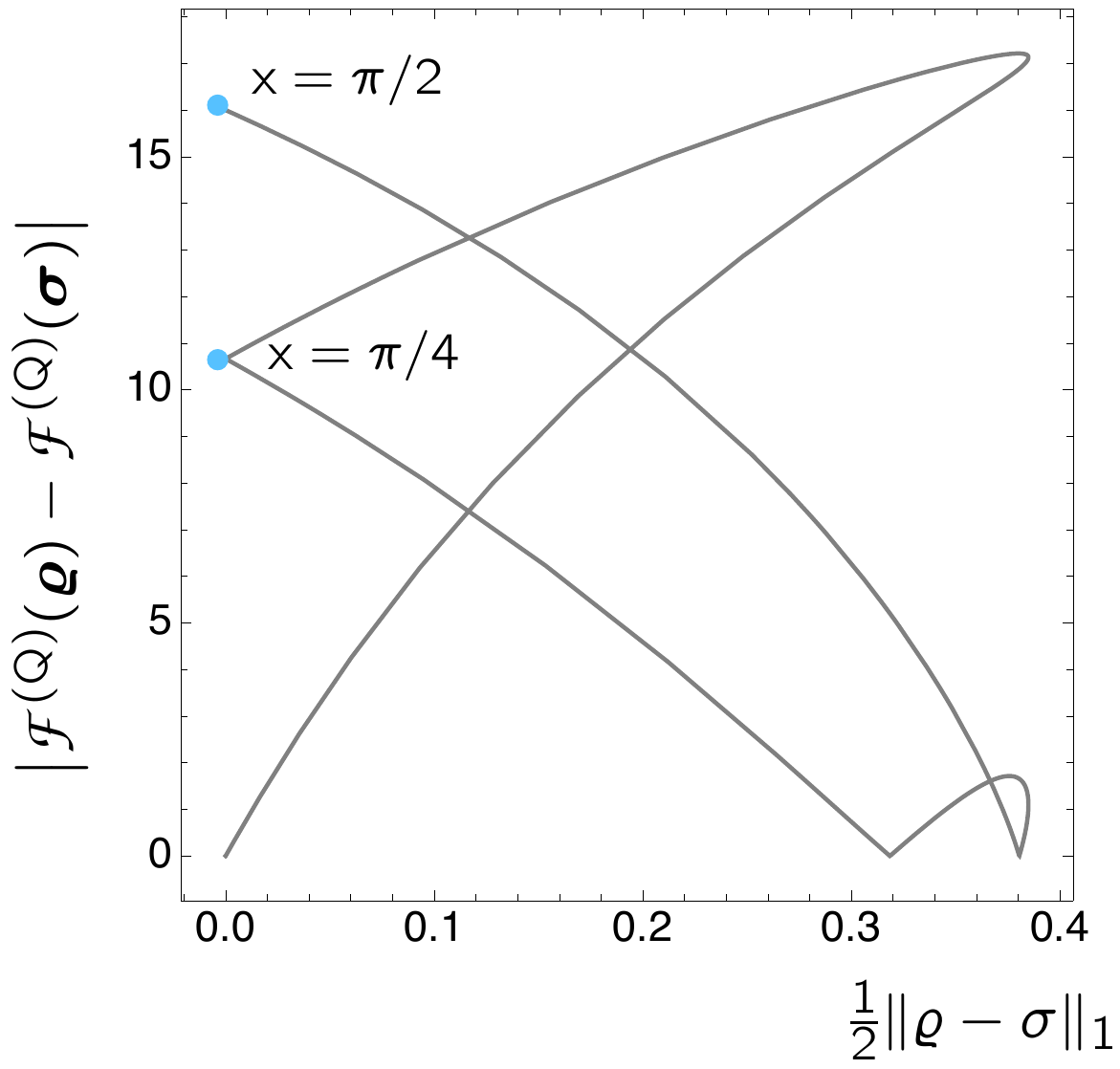}\\ \vskip4mm
\includegraphics[scale=0.37]{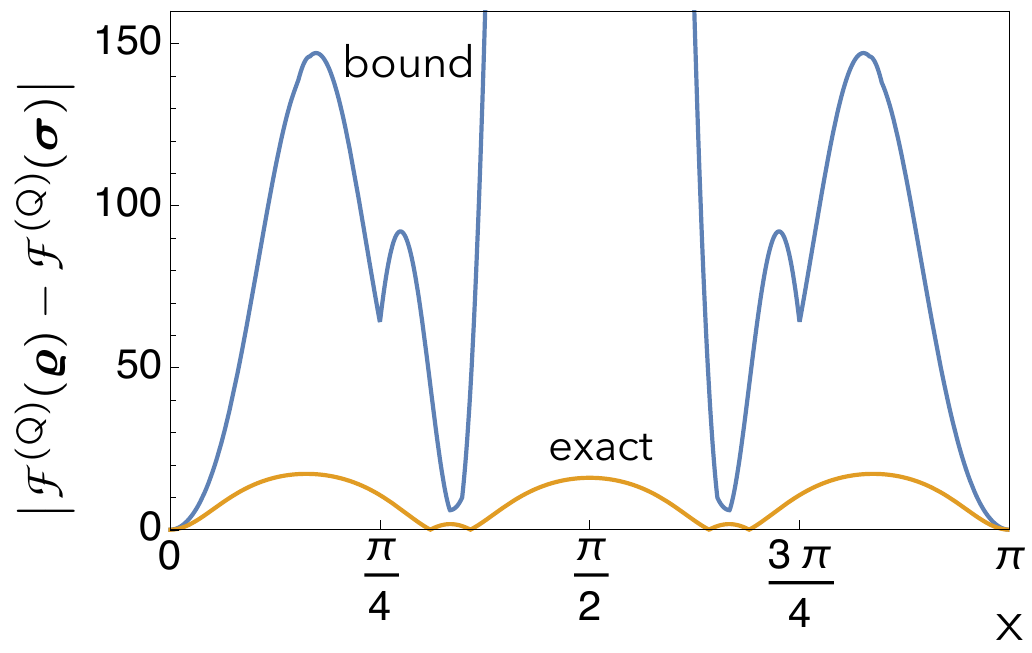}
\caption{(Top): Difference of the QFIs for two density matrices given in subsec. \ref{Qubit} vs. their distance, for $ x \in [0,\pi/2]$. It is evident, from the multivaluedness of $\big|\mathpzc{F}^{(\mathrm{Q})}(\bm{\varrho})- \mathpzc{F}^{(\mathrm{Q})}(\bm{\sigma})\big|$ at the point where $\Vert \varrho-\sigma \Vert_1=0$, that at $ x =\pi/4, \pi/2$ the QFI exhibits violation of the reduced continuity (while satisfying the continuity in the general sense of Theorem \ref{thm:1}). (Bottom): Comparison of the exact value of $\big|\mathpzc{F}^{(\mathrm{Q})}(\bm{\varrho}) - \mathpzc{F}^{(\mathrm{Q})}(\bm{\sigma})\big|$ and our upper bound (\ref{continuity-2}).}
\label{fig:fish-dist}
\end{figure}

Here we show through a simple example that the difference of the QFIs can depend on both the distance of the states and the distance of their derivatives. To this end, we consider a one-qubit state ($N=1$) in the form of
\begin{equation}
\varrho(x) = \big(\mathbbmss{I}+r_{\varrho}(x)S_z\big)/2,
\label{diagonal}
\end{equation}
represented in the computational basis $\{|0\rangle,|1\rangle\}$, where $S_z=|0\rangle\langle 0|-|1\rangle\langle 1|=\mathrm{diag}(1,-1)$ is the $z$-Pauli matrix. Since the state is diagonal and full-rank (perhaps except few points), the SLD can be readily calculated as $L_{\bm{\varrho}}=[\mosh r_{\varrho}/(1+r_{\varrho})]|0\rangle\langle 0|-[\mosh r_{\varrho}/(1-r_{\varrho})]|1\rangle\langle 1|$. Thus, the QFI reads as
\begin{equation}
\mathpzc{F}^{(\mathrm{Q})}(\bm{\varrho})=(\mosh r_{\varrho})^2/(1-r_{\varrho}^2).
\end{equation}
Taking $\varrho$ with $r_{\varrho}=\sin^2(x)$ and $\sigma$ with $r_{\sigma}=\sin^2( 3 x )$, it is straightforward to see that at $ x =\pi/4$, $\varrho=\sigma$, whereas $\mathpzc{F}^{(\mathrm{Q})}(\bm{\varrho})\neq \mathpzc{F}^{(\mathrm{Q})}(\bm{\sigma})$. To illustrate this result, $|\mathpzc{F}^{(\mathrm{Q})}(\bm{\varrho})-\mathpzc{F}^{(\mathrm{Q})}(\bm{\sigma})|$ has been depicted in terms of $\Vert \varrho-\sigma \Vert_1/2=|r_{\varrho}-r_{\sigma}|/2$ in Fig. \ref{fig:fish-dist} (top). It is seen that at $ x =0$ the two states and their associated QFIs are equal due to the equality of the derivatives of the states there, $\mosh\varrho(0)=\mosh\sigma(0)=0$; whereas at $ x =\pi/4, \pi/2$ the QFI exhibits a diversion from the \textit{reduced} continuity. We have also compared our bound and the exact value of the difference of the QFIs (vs. $x$) in Fig. \ref{fig:fish-dist} (bottom). Note that our bound at point $ x =\pi/2$ (where $P_{\varrho}$ is not differentiable because its rank changes) becomes trivial. In fact, the divergence of our bound at $ x =\pi/2$ can be a case of the failure of our continuity relation---see remark (iv) of the previous section.

\textit{Remark.}---Because of the diagonal form of the density matrices (\ref{diagonal}), this example can be alternatively explained by the continuity relation for the classical Fisher information [Eq. (\ref{ext-CFI})] for probability distributions $(1/2)\{1+\sin^{2}(x),\cos^{2}(x)\}$ and $(1/2)\{1+\sin^{2}(3 x ),\cos^{2}(3 x )\}$.

\subsection{Exponential density matrices}

As another example, we consider states in the exponential form \cite{Jiang}
\begin{align}
\varrho (x)=e^{H(x)},
\sigma (x)=e^{G(x)}.
\end{align}
This class includes, for example, thermal states. We have $\mosh\varrho =\int _{0}^{1}\mathrm{d}s~e^{sH}~\mosh H~e^{(1-s)H}$. Hence, Eq. (\ref{new-ineq}) and $ \Vert e^{sH}\Vert_{\infty}=\Vert \varrho^{s}\Vert_{\infty}\leqslant 1$ (for $0\leqslant s\leqslant 1$) yield
\begin{align}
\Vert \mosh\varrho\Vert_{1}\leqslant\Vert \mosh H\Vert_{1}  ,\Vert \mosh\sigma\Vert_{1}\leqslant \Vert \mosh G\Vert_{1}.
\end{align}
After some algebra one can also show that (see appendix \ref{app:eq-proof})
\begin{align}
\Vert \varrho - \sigma\Vert_{1}\leqslant & \Vert H-G\Vert_{1},\nonumber\\
\Vert \mosh\varrho - \mosh\sigma\Vert_{1} \leqslant & \frac{1}{2}\Vert H-G \Vert_{\infty} \big(\Vert\mosh H\Vert_{1} + \Vert\mosh G \Vert _{1}\big) \nonumber\\
&+\Vert\mosh H-\mosh G \Vert_{1}.
\label{mosh-exp-dif}
\end{align}

In this case, $ \varrho$ and $ \sigma $ are full-rank density matrices, thus the absolute value of the difference of their QFIs is given by Eq. (\ref{continuity-2}) as
\begin{align}
\big|\mathpzc{F}^{(\mathrm{Q})}(\bm{\varrho})-\mathpzc{F}^{(\mathrm{Q})}(\bm{\sigma})\big| \leqslant& f^{(\mathrm{Q})}_{\mathrm{e}}\Vert H-G\Vert_{1}+ g^{(\mathrm{Q})}_{\mathrm{e}} \Vert\mosh H-\mosh G\Vert_{1} ,
\label{continuity-exp}
\end{align}
where
\begin{align}
f^{(\mathrm{Q})}_{\mathrm{e}} & =\frac{\Vert\mosh G\Vert_{1} ^{2}}{e^{\lambda_{\min}(H)+\lambda_{\min}(G) }}+\frac{\left( \Vert\mosh H\Vert_{1}+\Vert\mosh G\Vert_{1} \right) ^{2}}{2e^{\lambda_{\min}(H) }}\nonumber\\
g^{(\mathrm{Q})}_{\mathrm{e}} & =\frac{1}{e^{\lambda_{\min}(H)}}\big( \Vert\mosh H\Vert_{1}+\Vert\mosh G\Vert_{1}\big) .
\end{align}

For the special case where the dependence on the unknown parameter is linear, i.e., $H(x)= x \, H_{1}$ and $G(x)= x \, G_{1}$, Eq. (\ref{continuity-exp}) becomes the reduced continuity relation. An example of this case is thermometry, where $x$ is the (minus) inverse temperature and $H_{1}$ and $G_{1}$ are Hamiltonians \cite{Correa-et-al}.

\begin{figure}[tp]
\includegraphics[scale=.35]{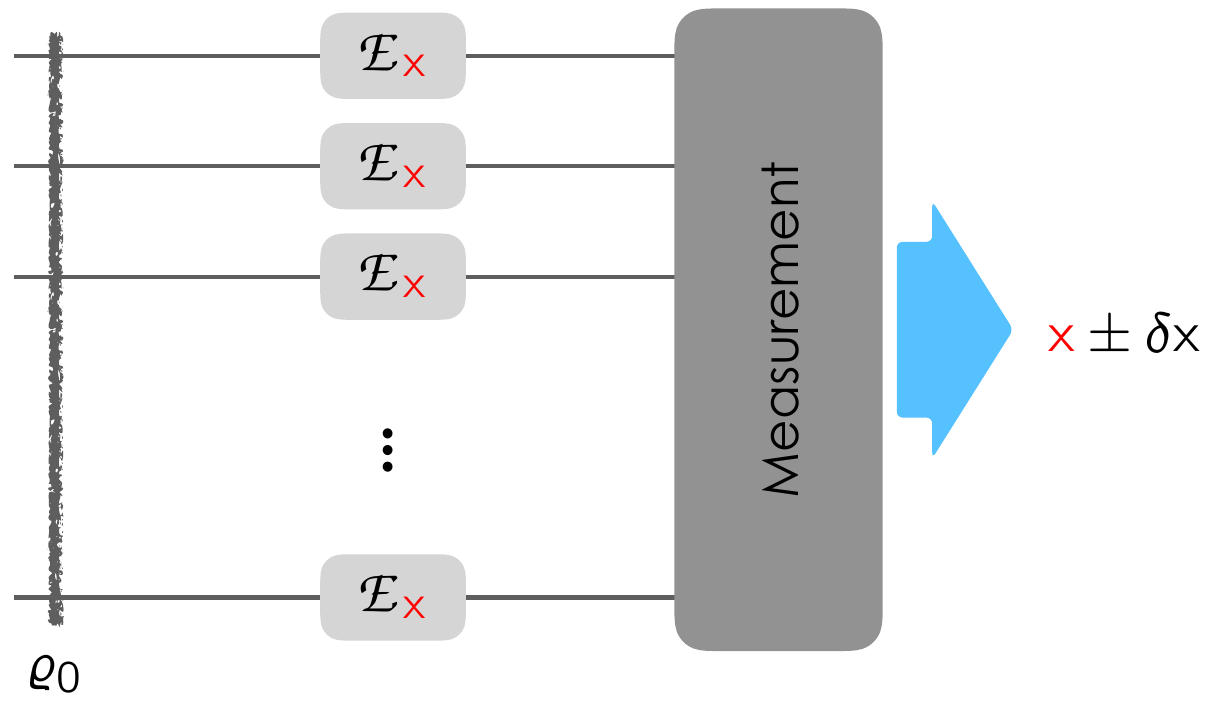}
\caption{Schematic of a metrological scenario with general parameter-dependent dynamics represented by a linear map $\mathpzc{E}_{x}$. Note that this is an example in line with the general scheme of Fig. \ref{fig:fig-1} (top).}
\label{fig:gen}
\end{figure}

\subsection{General noiseless quantum dynamics: Parameter encoding by a noiseless quantum channel}
\label{example:noiseless}

We recall that in Fig. \ref{fig:fig-1} we portrayed a generic open-system \textit{dynamical} scenario where preparation and dynamics (and perhaps measurement) may be affected by further uncontrollable noise. Here we want to partially relax this generality in the sense that we assume only the preparation is affected by noise whilst the dynamics is intact and ideal. We show that in this reduced case, the reduced continuity is relevant. 

Consider a (parameter-dependent) dynamical map $\mathpzc{E}_{x}$ with the Kraus representation $\mathpzc{E}_{x}[\circ]=\sum_{k=1}^{q} A_k \circ A_k^{\dag}$ \cite{book:Hayashi,book:Nielsen,book:Watrous}, in which we have dropped the explicit dependence of $A_{k}$s on $x$ in order to avoid cluttering the notation. Applying $N$ identical and independent maps $\mathpzc{E}_{x}^{\otimes N}$ on $N$-particle initial probe states (see Fig. \ref{fig:gen}) gives
\begin{align}
\mathpzc{E}_{x}^{\otimes N}[\varrho_{0}]:=\sum_{\mathbf{k}
}A_{\mathbf{k}}^{(N)}\varrho_{0}A_{\mathbf{k}}^{(N)\dagger},
\label{Kraus-rep}
\end{align}
where $A_{\mathbf{k}}^{(N)}=A_{k_1}\otimes A_{k_2}\otimes\cdots\otimes A_{k_N}$, with $\mathbf{k}=(k_1,k_2,\ldots,k_N)$. Here $k_j\in \{1,2,\ldots,q\}$ is the $k_j$th Kraus operator for the dynamics of the $j$th probe system (subscript $j\in \{1,2,\ldots,N\}$ runs over system numbers \cite{Fujiwara}). 

\begin{figure}[tp]
\includegraphics[scale=.2]{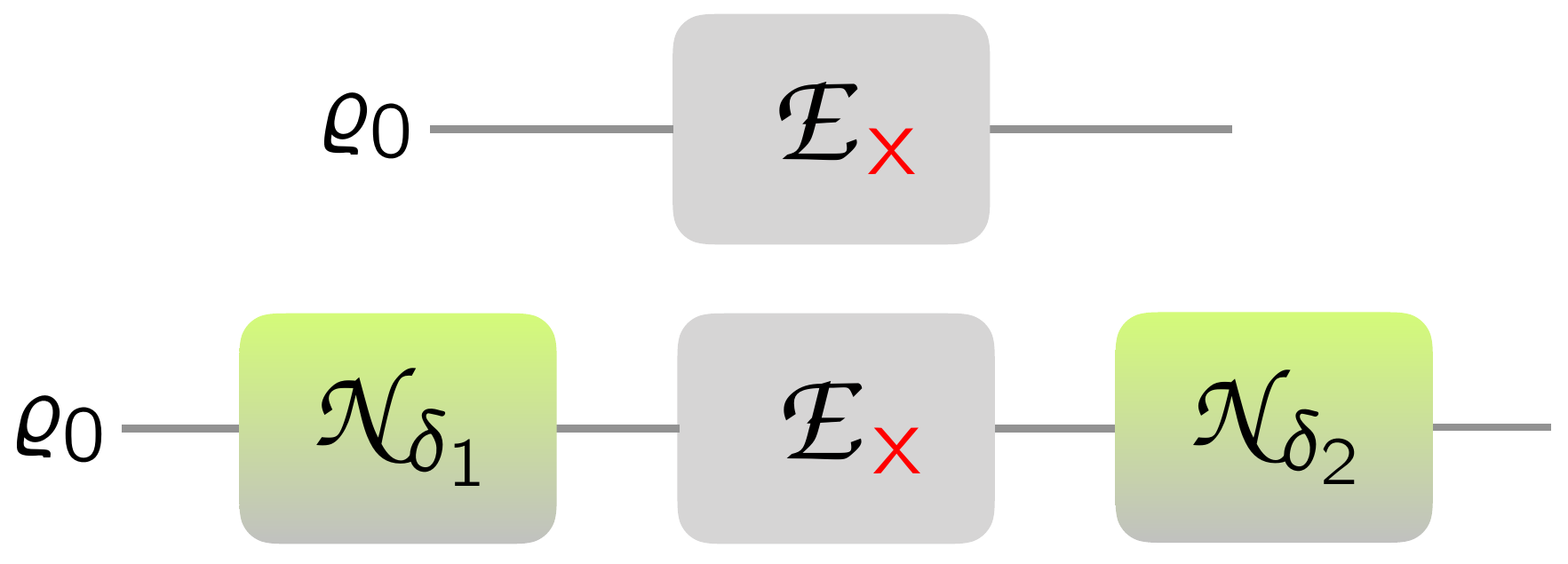}
\caption{Schematic of a metrological scenario with general parameter-dependent dynamics $\mathpzc{E}_{x}$. (Top): The noiseless scenario. (Bottom): The noisy scenario. Here the noise operations $\mathpzc{N}_{\,\delta_{1}}$ and $\mathpzc{N}_{\,\delta_{2}}$ affect, respectively, the preparation and the dynamics.}
\label{fig:noise}
\end{figure}

Considering the dynamics of Eq. (\ref{Kraus-rep}) applied on the two $N$-particle initial states $\varrho_0$ and $\sigma_0$, one obtains
\begin{align}
\label{aval}
&\Vert\mosh\varrho-\mosh\sigma\Vert_{1}
=\\
&\Big\Vert\sum_{\mathbf{k}}\mosh A_{\mathbf{k}}^{(N)}(\varrho_{0}
-\sigma_{0})A_{\mathbf{k}}^{(N)\dagger}+A_{\mathbf{k}}^{(N)}(\varrho_{0}
-\sigma_{0})\mosh A_{\mathbf{k}}^{(N)\dagger}\Big\Vert_{1}. \nonumber
\end{align}
Using the triangle inequality and Eq. (\ref{new-ineq}) yields
\begin{align}
&\Vert\varrho-\sigma\Vert_{1}\leqslant \Vert\varrho_{0}-\sigma_{0}\Vert_{1}\sum_{\mathbf{k}}
\big\Vert A_{\mathbf{k}}^{(N)}\big\Vert_{\infty} ^{2}, \label{dif-gen-dyn}\\
&\Vert\mosh\varrho-\mosh\sigma\Vert_{1}\leqslant 2\Vert\varrho_{0}-\sigma_{0}\Vert_{1} \sum_{\mathbf{k}}\big\Vert\mosh A_{\mathbf{k}}^{(N)}\big\Vert_{\infty}\, \big\Vert A_{\mathbf{k}}^{(N)}\big\Vert_{\infty}.
\label{mosh-gen-dyn}
\end{align}

These imply that for any pair of states $ \varrho $ and $ \sigma $ obtained from the same dynamics, the continuity of the QFI in the general form (\ref{continuity-2}) is simplified to the reduced continuity,
\begin{equation}
\big|\mathpzc{F}^{(\mathrm{Q})}(\bm{\varrho})-\mathpzc{F}^{(\mathrm{Q})}(\bm{\sigma})\big| \leqslant h(\varrho_{0},\sigma_{0},\mathpzc{E})\, \Vert \varrho_{0} -\sigma_{0}\Vert_{1},
\label{con-EE}
\end{equation}
where the explicit form of $h$ can be read from Eqs. (\ref{continuity-2}), (\ref{dif-gen-dyn}), and (\ref{mosh-gen-dyn}) as 
\begin{equation}
h=f^{(\mathrm{Q})} \sum_{\mathbf{k}}\Vert A_{\mathbf{k}}^{(N)}\Vert_{\infty}^{2} + 2g^{(\mathrm{Q})} \sum_{\mathbf{k}}\Vert A_{\mathbf{k}}^{(N)}\Vert_{\infty}\, \Vert \partial_{x}A_{\mathbf{k}}^{(N)}\Vert_{\infty},
\end{equation}
with $f^{(\mathrm{Q})}$ and $g^{(\mathrm{Q})}$ defined in Eqs. (\ref{def:f}) and (\ref{def:g}). We can further simplify $h$ if we use the relation between $\lambda_{\min}(\mathpzc{E}_{x}[\varrho_{0}])$ and $\lambda_{\min}(\varrho_{0})$ proven in appendix \ref{app:eig}. The advantage of this relation over the result of Ref. \cite{Kolod} is two-fold: (i) the dynamics here is a general quantum channel, not limited to the unitary evolutions, and (ii) dependence of the dynamics on the parameter is arbitrary (but differentiable), not necessarily linear. In the case of the unitary evolution $ U_{x}=e^{-i x H} $, Corollary \ref{corr:2} gives
\begin{align}
\big|\mathpzc{F}^{(\mathrm{Q})}(\bm{\varrho})-\mathpzc{F}^{(\mathrm{Q})}(\bm{\sigma})\big|
\leqslant  \frac{4\Vert H\Vert_{\infty} ^{2} \big(2+\lambda_{\min}(\sigma_{0})\big)}{\lambda_{\min}(\varrho_{0}) \, \lambda_{\min}(\sigma_{0})}\Vert\varrho_{0} -\sigma_{0}\Vert_{1}.
\label{con-unitary}
\end{align}
One can compare this with the bound reported in Ref. \cite{Kolod},
\begin{align}
\big|\mathpzc{F}^{(\mathrm{Q})}(\bm{\varrho})-\mathpzc{F}^{(\mathrm{Q})}(\bm{\sigma})\big|
\leqslant  32\Vert H\Vert_{\infty} ^{2} \sqrt{\Vert\varrho_{0} -\sigma_{0}\Vert_{1}}.
\label{con-unitary-kolod}
\end{align}
It is important to highlight a particular advantage of the continuity relation (\ref{con-EE})---or similarly Eqs. (\ref{con-unitary}) and (\ref{con-unitary-kolod}). If we are given an initial state for a metrology scenario as described in this subsection, then by choosing $\sigma$ a ``simple'' state whose QFI one can calculate readily, we can find an estimate on the QFI of the evolved state $\varrho$ and depending on how it scales vs. $N$ we may be able to decide whether the initial state $\varrho_{0}$ is useful for metrology. This approach can give a significant computational advantage because typically calculating the QFI for manybody states is a formidable task---even numerically.

\subsection{Noisy quantum dynamics: Parameter encoding by a noisy quantum channel}
\label{example:noise}

Here we discuss an example of an open-system metrology scenario, depicted in Fig. \ref{fig:noise}, which is affected by noise in both preparation and dynamics steps (in line with the general scheme of Fig. \ref{fig:fig-1} (bottom)). Let $U(x)= (1/\sqrt{2}) (e^{i x S_{z}} -i S_{y})$, where $S_{y}$ is the $y$-Pauli matrix. This can describe the dynamics of a spin-$1/2$ particle under an external magnetic field $\mathbf{B}(x)$ which nonlinearly depends on an unknown parameter $x$ as
\begin{equation}
\mathbf{B}(x)=\frac{2\cos^{-1}\left[(1/\sqrt{2})\cos  x \right]}{\sqrt{1+\sin^{2} x }} \left(0,1, -\sin x  \right).
\end{equation}
That is, $U(x)=e^{-i H(x)}$, where $H(x)=-\mathbf{B}(x)\cdot \mathbf{S}$ and $\mathbf{S} =(1/2)(S_{x},S_{y},S_{z})$ is the vector of the Pauli operators (up to a factor of $1/2$). Here again we have taken $N=1$.
\begin{figure}[tp]
\includegraphics[scale=.32]{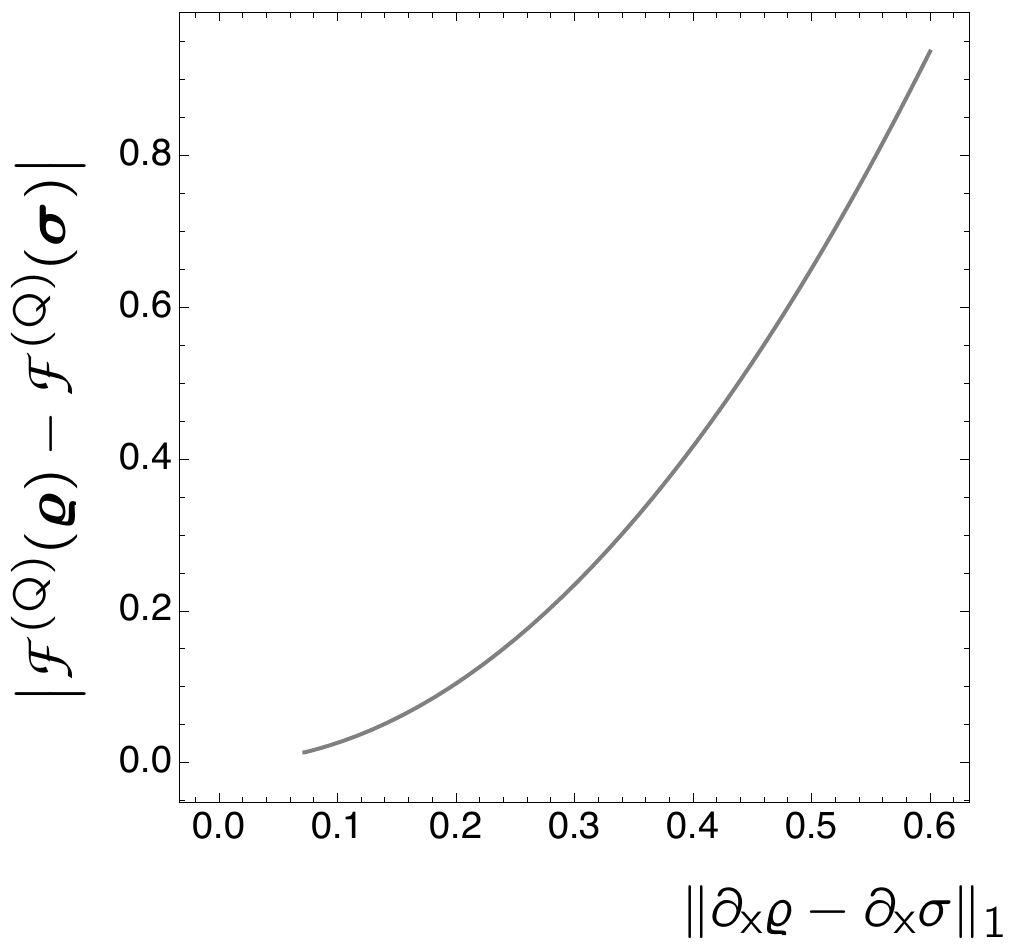}
\caption{Example \ref{example:noise}. $\big| \mathpzc{F}^{(\mathrm{Q})}(\bm{\varrho}) - \mathpzc{F}^{(\mathrm{Q})}(\bm{\sigma}) \big|$ vs. $\Vert \partial_{x}\varrho - \partial_{x}\sigma\Vert_{1}$. This monotonic behavior indicates that the very nonvanishing of the difference of the QFIs is rooted at the nonvanishing property of $\Vert \partial_{x}\varrho - \partial_{x}\sigma \Vert_{1}$.}
\label{fig:diff-last}
\end{figure}

\begin{figure*}[tp]
\includegraphics[scale=.29]{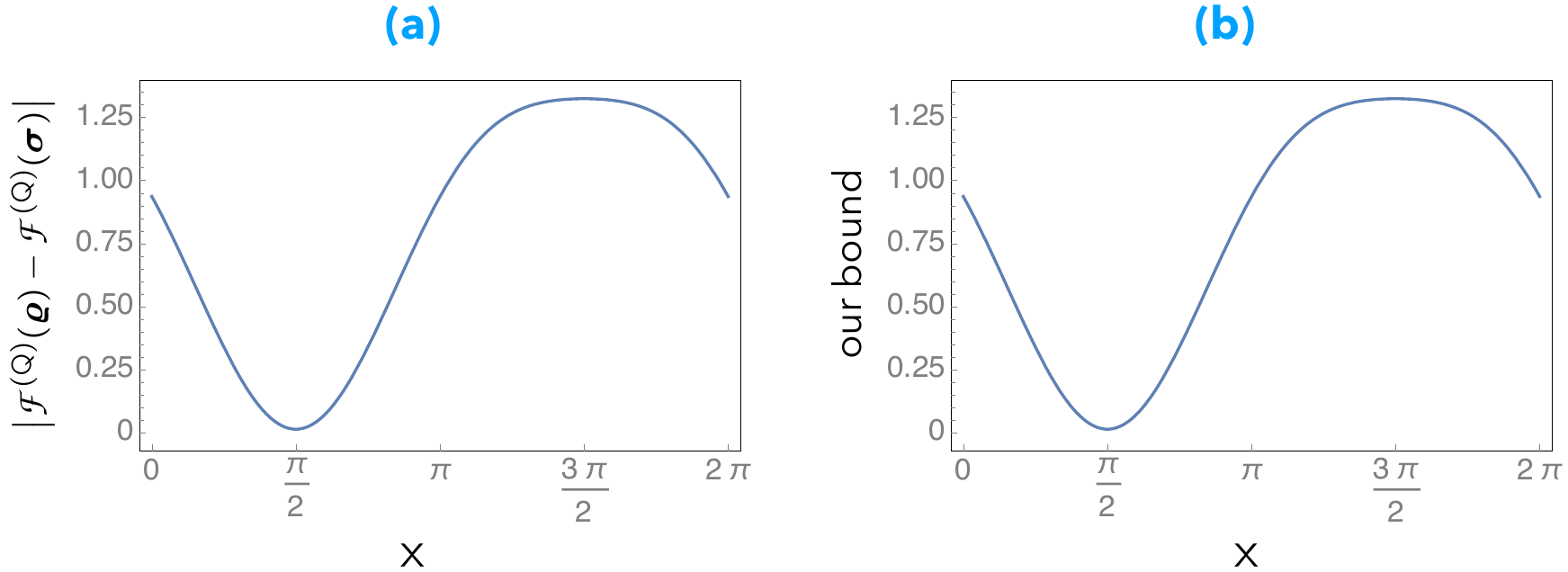} \hskip6mm \includegraphics[scale=.29]{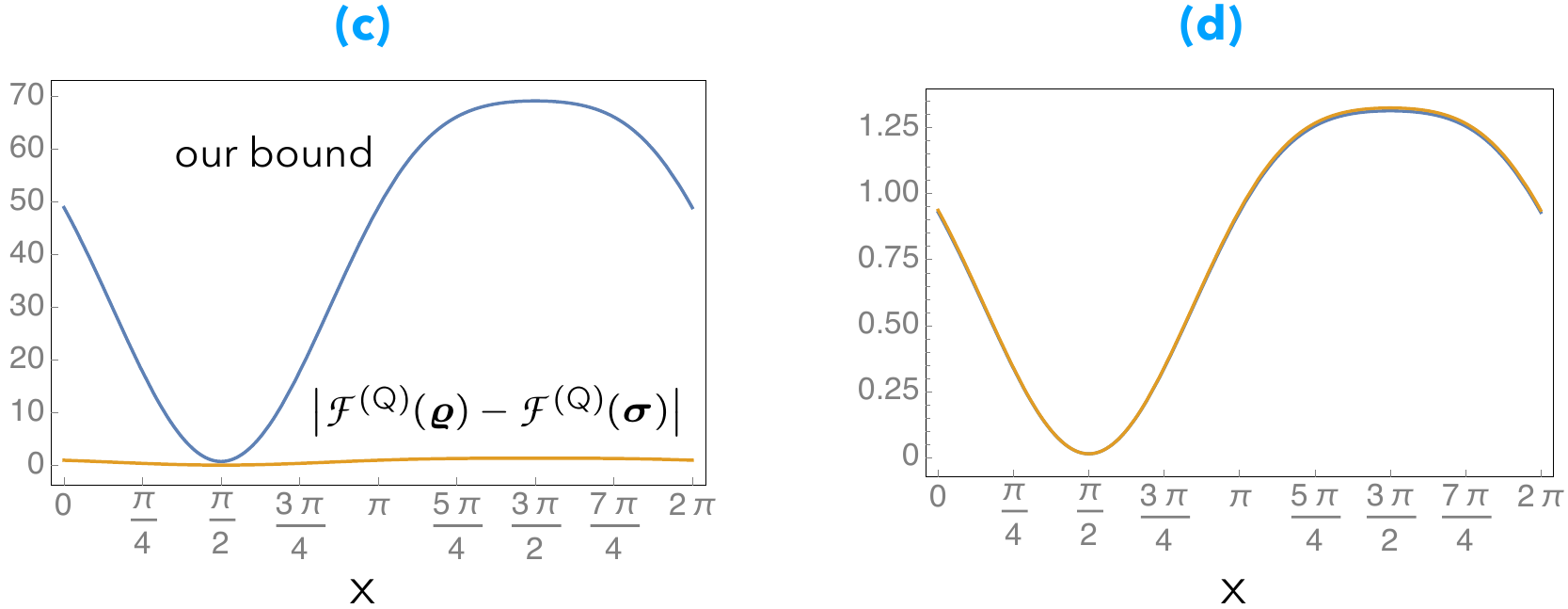}
\caption{Example \ref{example:noise}. (a): Exact value of $\big| \mathpzc{F}^{(\mathrm{Q})}(\bm{\varrho}) - \mathpzc{F}^{(\mathrm{Q})}(\bm{\sigma}) \big|$ vs. $x$. (b): Bound (\ref{continuity-2}) vs. $x$. (c): The two quantities in the same plot. (d) The two quantities in the same plot where our bound is rescaled with a factor of $\approx 0.02$. The good agreement here indicates that our bound captures the behavior of the difference of the exact QFIs faithfully.}
\label{fig:noisy-general}
\end{figure*}

The goal now is to estimate $x$. Here $\mathpzc{E}_{x}[\circ] = U(x) \circ U^{\dag}(x)$. We assume that this operation is now affected by two depolarizing noise channels---defined by $\mathpzc{N}_{\,\delta}[\xi] = (1-\delta) \xi +(\delta/2)\mathbbmss{I}$ for any state $\xi$---and is modified to $\mathpzc{E}'_{x} = \mathpzc{N}_{\,\delta_{1}}\mathpzc{E}_{x}\mathpzc{N}_{\,\delta_{2}}$. This scheme can model noise in preparation and dynamics scenarios. The noise parameters $\delta_{1}$ and $\delta_{2}$ are taken to be independent of $x$. For specificity, we assume $\varrho_{0}=(1/2)(\mathbbmss{I}+ \mathbf{r}_{0}\cdot\mathbf{S})$, where $\mathbf{r}_{0}=(0,-1/\sqrt{3},1/\sqrt{2})$. Figure \ref{fig:noisy-general} shows the exact values of $\big| \mathpzc{F}^{(\mathrm{Q})}(\bm{\varrho}) - \mathpzc{F}^{(\mathrm{Q})}(\bm{\sigma}) \big|$ and our bound (\ref{continuity-2}), where $\varrho=\mathpzc{E}_{x}[\varrho_{0}]$ and $\sigma=\mathpzc{E}'_{x}[\varrho_{0}]$ and we have assumed $\delta_{1}=\delta_{2}=1/3$. As is evident from this figure, our bound captures the behavior of the exact difference of the QFIs faithfully (see Fig. \ref{fig:noisy-general}-(d)); with a rescaling factor of $\approx 0.02$ the two quantities have almost similar values for all $x$'s. Interestingly, one can discern that the very nonvanishing of the difference of the QFIs is associated with the nonvanishing property of $\Vert \partial_{x}\varrho - \partial_{x}\sigma \Vert_{1}$---Fig. \ref{fig:diff-last}. Note that for this example, $\Vert \varrho -\sigma\Vert_{1} = \sqrt{5/6}(\delta_{1}+ \delta_{2} - \delta_{1}\delta_{2})$, independent of $x$; whereas $\Vert \partial_{x}\varrho - \partial_{x}\sigma\Vert_{1} = \sqrt{1/6}(\delta_{1}+ \delta_{2} - \delta_{1}\delta_{2})\sqrt{6+\cos 2 x -2\sqrt{6}\sin  x }$.

\section{Summary}
\label{sec:summary}

We have proved the continuity relations for the quantum Fisher information (QFI) and the symmetric logarithmic derivative (SLD), which was motivated by the observation that the classical Fisher information features a somewhat akin fundamental property. These properties imply that, in general cases, the QFI and the SLD behave such that for two close states with close first derivatives both the QFIs and the SLDs would be respectively close too. These continuity relations are completely general in various aspects: (i) They are irrespective of dynamics or how the parameter dependence enters the description of the states, thus they are applicable to any metrology scenarios. (ii) They hold for any dependence of the states on the unknown parameter (modulo differentiability). (iii) They hold for any state whether full-rank or incomplete-rank. To establish the latter feature, we have introduced a regularized-SLD by removing the inherent singularity of the SLD for vanishing eigenvalues of the state density matrix. Notwithstanding this generality, our proofs are fairly straightforward, based mostly on well-known operator norm inequalities.

In addition, we have demonstrated that in the special case where the dependence of the states on the unknown parameter is induced by a quantum channel, the continuity holds in its reduced form, i.e., only with respect to the initial states. Nevertheless, for the case when one initial state evolves through two different quantum channels, we have shown that the continuity relation should be considered in its general form. The latter situation includes open-system metrology where one of the maps represents ideal dynamics whereas the other map represents the real (noisy) dynamics.

We anticipate that, given the generality and utility of our continuity relation for the QFI, it can spur numerous applications in quantum metrology and other areas of quantum information science and technology. For example, it may help significantly reduce computational cost of deciding whether a given initial state is useful for a quantum metrology task, by obviating the need to compute the QFI explicitly. This application can be important in quantum sensing.

\textit{Acknowledgements}.---This work was partially supported by Sharif University of Technology's Office of Vice President for Research and Technology through Grant No. QA960512, the School of Nano Science at the Institute for Research in Fundamental Sciences (IPM), and the Academy of Finland's Center of Excellence program QTF Project 312298.


\vfill \pagebreak \onecolumngrid
\appendix

\section{Proof of the continuity relation for the classical Fisher information}
\label{app:CFI-proof}

Here we show that the classical Fisher information (\ref{CFI-text}) is not necessarily continuous in the ordinary or \textit{reduced} sense (i.e., in the sense $| p(y|x) - q(y|x)|\to0$ then $\big|\mathpzc{F}^{(\mathrm{C})}(\{p\})-\mathpzc{F}^{(\mathrm{C})}(\{q\})\big|\to0$). Rather, it fulfills a continuity relation which includes both expressions $| p(y|x) - q(y|x)|$ and $| \partial_{x}p(y|x) - \partial_{x}q(y|x)|$. 

For two different conditional probability distributions $p(y|x)$ and $q(y|x)$, associated to probability distributions $\{p\}$ and $\{q\}$ for the same unknown parameter $x$, we write 
\begin{align}
\mosh p(y|x)&=p(y|x) \,\ell _{p},\label{def-l}\\
\mosh q(y|x)&=q(y|x) \, \ell_{q},\label{def-lp}
\end{align}
where $\ell_{p}=\partial_{x}\log p(y|x)$ and $\ell_{q}=\partial_{x}\log q(y|x)$. By starting from the definition, we have
\begin{align}
\Big\vert\mathpzc{F}^{(\mathrm{C})}(\{p\})-\mathpzc{F}^{(\mathrm{C})}(\{q\})\Big\vert =& \Big\vert\int_{\mathpzc{D}_{ y }}\mathrm{d} y \, \Big(\mosh p(y|x)\ell_{p}-\mosh q(y|x)\ell_{q}\Big)\Big\vert\nonumber\\
 \leqslant& \int_{\mathpzc{D}_{ y }}\mathrm{d} y \,\Big(\big\vert \mosh p(y|x)-\mosh q(y|x)\big\vert \,\vert\ell_{p}\vert +\big\vert\mosh q(y|x)\big\vert \,\vert\ell_{p}-\ell_{q}\vert\Big),
\label{dif-CFI}
\end{align}
where in the last line we have employed the triangle inequality and 
\begin{equation}
|AB-A'B'|\leqslant |A-A'|\,|B| + |A'|\,|B-B'|.
\label{abs-diff}
\end{equation}
In addition, it is straightforward to show that 
\begin{align}
\label{l-lp}
\vert\ell_{p}-\ell_{q}\vert\leqslant&\frac{1}{\vert p(y|x)\,q(y|x)\vert}\Big( \vert \mosh p(y|x) \, \mosh q(y|x) \vert\, \vert q(y|x)\vert  + \vert \mosh q(y|x) \vert \,\vert p(y|x)-q(y|x)\vert\Big).
\end{align}
Now Eqs. (\ref{dif-CFI}) and (\ref{l-lp}) yield Eq. (\ref{ext-CFI}), 
\begin{align}
\Big\vert\mathpzc{F}^{(\mathrm{C})}[p(y|x)]-\mathpzc{F}^{(\mathrm{C})}[q(y|x)]\Big\vert 
&\leqslant f^{(\mathrm{C})} \int_{\mathpzc{D}_{ y }}\mathrm{d}y\, \big\vert p(y|x)-q(y|x)\big\vert + g^{(\mathrm{C})} \int_{\mathpzc{D}_{ y }}\mathrm{d}y\, \big\vert \mosh p(y|x)-\mosh q(y|x)\big\vert ,
\end{align}
with 
\begin{align}
f^{(\mathrm{C})}&=\max_{ y \in\mathpzc{D}_{ y }}\frac{\vert\mosh q(y|x)\vert ^{2}}{\vert p(y|x)\, q(y|x)\vert}, \label{f-class}\\
g^{(\mathrm{C})}&=\max_{ y \in\mathpzc{D}_{ y }}\frac{1}{\vert p(y|x)\vert}\big(\vert\mosh p(y|x)\vert +\vert\mosh q(y|x)\vert \big). \label{g-class}
\end{align}
\hfill$\square$

It is interesting that despite simplicity of this relation, to the best of our knowledge it had never been conceived earlier in the literature.

\section{The SLD [Eq. (\ref{eq:SLD-int})] and r-SLD [Eq. (\ref{sing-L-2})]}
\label{app:SLD}

Equation (\ref{eq:SLD}) is reminiscent of the Lyapunov equation \cite{Paris:tut,book:Hayashi}, which in turn is a special case of the Sylvester equation \cite{book:Bhatia},
\begin{equation}\label{Sylvester}
AX-XB=Y.
\end{equation}
From Theorem 7.2.3 in Ref. \cite{book:Bhatia}, if $A$ and $B$ have disjoint spectrums, then
\begin{equation}\label{Syl-int}
X=\int_{0}^{\infty}{\mathrm{d}s\, e^{-sA}Ye^{sB}}.
\end{equation}
In our case of interest, we see that by taking $ X=L_{\bm{\varrho}} $, $ Y=2\mosh\varrho$, and $ A=-B=\varrho$, Eq. (\ref{eq:SLD}) has an integral  representation as in Eq. (\ref{Syl-int}) when the density matrix is full-rank. Below, we follow a careful analysis to examine utility of this integral representation for the SLD.

Suppose that $\varrho(x)$ is defined on a Hilbert space $\mathpzc{H}^{\otimes N}$ (where we keep, for a while, the $x$-dependence for clarity and to remind that the rank of $\varrho(x)$ may vary with $x$). In order to retain physical relevance of $\varrho(x)$, we assume sufficient smoothness for it in terms of $x$. We denote the eigenvectors of $\varrho(x)$ with $\{|\lambda_{i}(x)\rangle\}_{i=1}^{\mathpzc{H}^{\otimes N}}$, and more specifically assume $\{\ket{s_{j}(x)}\}$ and $\{\ket{n_{k}(x)}\}$ denote, respectively, those eigenvectors of $\varrho(x)$ which correspond to the nonzero  and zero eigenvalues, i.e., support vectors and null space vectors. Thus, for all $ x \in\mathpzc{D}_{x}$ we have 
\begin{align}
\varrho(x) &=\sum_{i =1}^{\mathpzc{H}^{\otimes N}} \lambda_{i}(x) |\lambda_{i}(x)\rangle \langle\lambda_{i} (x)|, 
\label{null-den}
\\
\mathbbmss{I}_{\mathpzc{H}^{\otimes N}} & = \sum_{i =1}^{\mathpzc{H}^{\otimes N}} |\lambda_{i}(x)\rangle \langle\lambda_{i} (x)|= P(x) + P_{\perp}(x), \label{Identity-x}
\end{align}
where $P(x) = \sum_{ |s_{j}\rangle \in\, \mathrm{supp}(x)} \ket{s_{j}(x)} \bra{s_{j}(x)}$ is the projection onto the support space of $\mathpzc{H}^{\otimes N}$. We obtain
\begin{equation}
\partial_{x}\varrho(x) = \sum_{i = 1}^{\mathpzc{H}^{\otimes N}}\partial_{x}\lambda_{i}(x) |\lambda_{i}(x)\rangle \langle\lambda_{i} (x)| +\sum_{ |s_{j}\rangle \in\, \mathrm{supp}(x)} \lambda_{j}(x) \big( \ket{\partial_{x} s_{j}(x)} \bra{s_{j}(x)} + \ket{s_{j}(x)} \bra{\partial_{x} s_{j}(x)}\big)
\end{equation}
and 
\begin{align}
e^{-s\varrho(x)} &=\sum_{i =1}^{\mathpzc{H}^{\otimes N}} e^{-s\lambda_{i}(x)} |\lambda_{i}(x)\rangle \langle\lambda_{i} (x)|=  \sum_{ |s_{j}\rangle \in\, \mathrm{supp}(x)} e^{-s\lambda_{j}(x)}\ket{s_{j}(x)} \bra{s_{j}(x)} + P_{\perp}(x), \label{e-to-}
\end{align}
where we identify the first term on the RHS as $e^{-s\widetilde{\varrho}(x)}$, with $\widetilde{\varrho}(x) = P(x) \varrho(x)P(x)$ being the restriction of $\varrho(x)$ onto its support space. From Eq. (\ref{e-to-}) the following relations are evident:
\begin{align}
 P_{\perp}(x) e^{-s\varrho(x)} =& e^{-s\varrho(x)} P_{\perp}(x) = P_{\perp}(x),\\
P_{\perp}(x) e^{-s\widetilde{\varrho}(x)} =& e^{-s \widetilde{\varrho}(x)} P_{\perp}(x) = 0,\\
e^{-s\varrho(x)} \partial_{x}\varrho(x) e^{-s\varrho(x)} =& e^{-s\widetilde{\varrho}(x)} \partial_{x}\varrho(x) e^{-s\widetilde{\varrho}(x)} + \sum_{ |s_{j}\rangle \in\, \mathrm{supp}(x)} \lambda_{j}(x) e^{-s\lambda_{j}(x)} \big( P_{\perp}(x)\ket{\partial_{x}s_{j}(x)} \bra{s_{j}(x)} + \ket{s_{j}(x)} \bra{\partial_{x}s_{j}(x)} P_{\perp}(x)\big)\nonumber\\
& + \sum_{|n_k\rangle \in \mathrm{null}(x)} \partial_{x}\lambda_{k}(x) |n_{k}(x)\rangle \langle n_{k} (x)|. \label{ere}
\end{align}
Note that the last term on the RHS of Eq. (\ref{ere}) is independent of $s$. By integrating the last equation over $s\in[0,t]$ and noting that $\int_{0}^{t}\mathrm{d}s\, e^{-s\lambda_{j}}=(1-e^{-t\lambda_{j}})/\lambda_{j}$ (for $\lambda_{j}\neq 0$), we obtain
\begin{align}
\int_{0}^{t} \mathrm{d}s\, e^{-s\varrho(x)} \partial_{x}\varrho(x) e^{-s\varrho(x)} =& \int_{0}^{t} \mathrm{d}s\,e^{-s\widetilde{\varrho}(x)} \partial_{x}\varrho(x) e^{-s\widetilde{\varrho}(x)} + \sum_{ |s_{j}\rangle \in\, \mathrm{supp}(x)} (1-e^{-t\lambda_{j}(x)}) \big( P_{\perp}(x)\ket{\partial_{x}s_{j}(x)} \bra{s_{j}(x)} + \ket{s_{j}(x)} \bra{\partial_{x}s_{j}(x)} P_{\perp}(x)\big) \nonumber\\
& + t\sum_{|n_k\rangle \in \mathrm{null}(x)} \partial_{x}\lambda_{k}(x) |n_{k}(x)\rangle \langle n_{k} (x)|.
\end{align}
Now multiplying this relation by $2$, taking the $t\to\infty$ limit, and recalling Eqs. (\ref{eq:SLD-int}) and (\ref{sing-L-2}) yield
\begin{align}
L_{\varrho(x)}= \mathpzc{L}_{\varrho(x)} + \sum_{|n_k\rangle \in \mathrm{null}(x)} \partial_{x}\lambda_{k}(x) |n_{k}(x)\rangle \langle n_{k} (x)|\times\lim_{t\to\infty}\int_{0}^{t}\mathrm{d}s.
\end{align}
The last term above vanishes identically in either of the following cases: (i) the rank of $\varrho(x)$ is always constant (whether full-rank or incomplete-rank) and does not depend on $x$, and (ii) $\varrho(x)$ is incomplete-rank but at our point of interest $x$ (where the rank changes) all vanishing eigenvalues have vanishing first derivatives too. In either of these cases the commonly-accepted integral form of the SLD and the r-SLD are equal, $L_{\varrho(x)}= \mathpzc{L}_{\varrho(x)}$. Otherwise, the above relation shows that the integral form (\ref{eq:SLD-int}) does not necessarily converge. Hence, the main advantage of the r-SLD is that, unlike the form in Eq. (\ref{eq:SLD-int}), it is always convergent and thus has a finite norm ($\Vert \mathpzc{L}_{\varrho(x)} \Vert_{\infty}<\infty$). In addition, this careful analysis can completely remove the confusion in the quantum metrology literature regarding the applicability of the integral representation (\ref{eq:SLD-int}) \cite{Wang,Saf}. This analysis also justifies the point we made in the main text that the r-SLD representation (\ref{sing-L-2}) is devoid of the divergence problem with the integral form (\ref{eq:SLD-int}), and thus this can make it more suitable. We, however, still need to justify that the r-SLD is relevant for the QFI. 

Despite the above discrepancy between the integral forms (\ref{eq:SLD-int}) and (\ref{sing-L-2}), we now show that the very basic definition of the SLD as in Eq. (\ref{eq:SLD}) always entails an indefiniteness for part of the SLD; the projection of the SLD on the null space of $\varrho$ ($\{\bra{n_{i}}L_{\bm{\varrho}}\ket{n_{j}}\}$) is left arbitrary. But interestingly, one can also show that this indefiniteness is completely irrelevant (i.e., does not contribute) as long as the QFI (\ref{def:qfi}) is concerned. Note that
\begin{align}
\mathpzc{F}^{(\mathrm{Q})}(\bm{\varrho})&=\mathrm{Tr}[L_{\bm{\varrho}}\varrho L_{\bm{\varrho}}]\nonumber\\
&=\sum_{|s_{i}\rangle\in\, \mathrm{support}}{\bra{s_{i}}L_{\bm{\varrho}}\varrho L_{\bm{\varrho}}\ket{s_{i}}}+\sum_{|n_{i}\rangle\in\, \mathrm{null}}{\bra{n_{i}}L_{\bm{\varrho}}\varrho L_{\bm{\varrho}}\ket{n_{i}}}\nonumber\\
&=\sum_{i}{\bra{s_{i}}L_{\bm{\varrho}}\,\mathbbmss{I}\,\varrho\,\mathbbmss{I}\, L_{\bm{\varrho}}\ket{s_{i}}}+\sum_{i}{\bra{n_{i}}L_{\bm{\varrho}}\, \mathbbmss{I} \,\varrho\,\mathbbmss{I}\, L_{\bm{\varrho}}\ket{n_{i}}}\nonumber\\
&=\sum_{i}{\bra{s_{i}}L_{\bm{\varrho}}\Big(\sum_{i_{1}}{\ket{s_{i_{1}}}\bra{s_{i_{1}}}+\ket{n_{i_{1}}}\bra{n_{i_{1}}}}\Big) \varrho \Big(\sum_{i_{2}}{\ket{s_{i_{2}}}\bra{s_{i_{2}}}+\ket{n_{i_{2}}}\bra{n_{i_{2}}}}\Big) L_{\bm{\varrho}}\ket{s_{i}}}\nonumber\\
&~~~~~+\sum_{i}{\bra{n_{i}}L_{\bm{\varrho}}\Big(\sum_{i_{2}}{\ket{s_{i_{3}}}\bra{s_{i_{3}}}+\ket{n_{i_{3}}}\bra{n_{i_{3}}}}\Big)\varrho\Big(\sum_{i_{3}}{\ket{s_{i_{4}}}\bra{s_{i_{4}}}+\ket{n_{i_{4}}}\bra{n_{i_{4}}}}\Big) L_{\bm{\varrho}}\ket{n_{i}}}\nonumber\\
&=\sum_{ii_{1}i_{2}}{\bra{s_{i}}L_{\bm{\varrho}}\ket{s_{i_{1}}}\bra{s_{i_{1}}}\varrho\ket{s_{i_{2}}}\bra{s_{i_{2}}} L_{\bm{\varrho}}\ket{s_{i}}}+\sum_{i,i_{3},i_{4}}{\bra{n_{i}}L_{\bm{\varrho}}\ket{s_{i_{3}}}\bra{s_{i_{3}}}\varrho\ket{s_{i_{4}}}\bra{s_{i_{4}}} L_{\bm{\varrho}}\ket{n_{i}}}\nonumber\\
&=\sum_{ij}\lambda_{j}\Big(\bra{s_{i}}L_{\bm{\varrho}}\ket{s_{j}}\bra{s_{j}} L_{\bm{\varrho}}\ket{s_{i}}+\bra{n_{i}}L_{\bm{\varrho}}\ket{s_{j}}\bra{s_{j}} L_{\bm{\varrho}}\ket{n_{i}}\Big),
\label{cal-QFI}
\end{align}
where no $\langle n_{i}|L_{\bm{\varrho}}|n_{j}\rangle$ term remains in the last expression.

\section{Bounds on eigenprojections}
\label{app:pp}

\begin{figure}[tp]
\includegraphics[scale=.4]{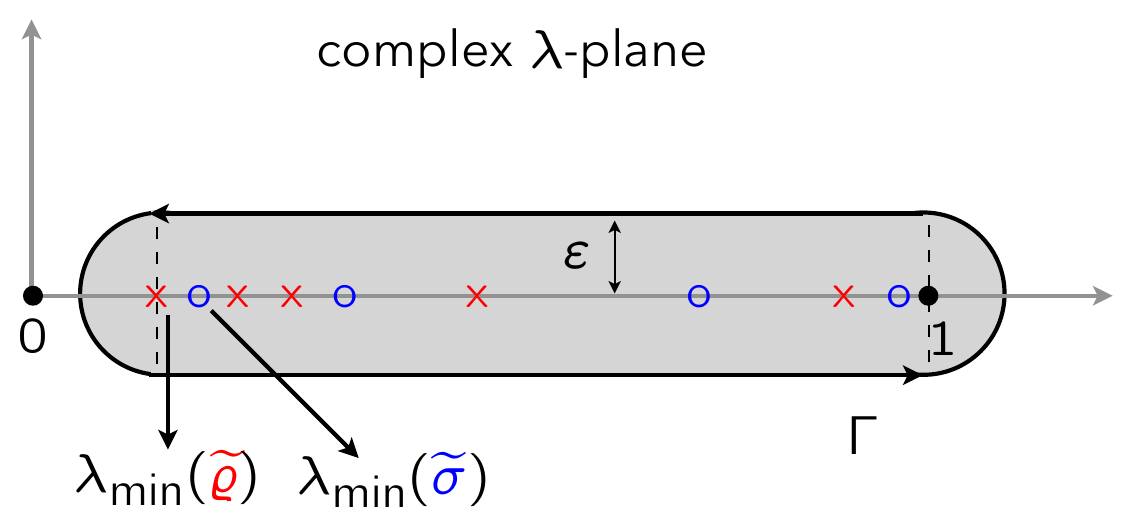}
\caption{A counter-clockwise integration contour in the form of a strip of width $2\varepsilon$, which encloses all nonvanishing eigenvalues of $\varrho$ and $\sigma$.}
\label{fig:boundary}
\end{figure}

\begin{lemma}
\label{lemma:1}
Let $P_{\varrho}$ and $P_{\sigma}$ denote eigenprojections on the support of two density matrices $\varrho$ and $\sigma$, respectively. Then,
\begin{align}
\Vert P_{\varrho}-P_{\sigma} \Vert_{\infty} &\leqslant \frac{8}{\lambda_{\min}(\widetilde{\varrho}) \,\lambda_{\min}(\widetilde{\sigma})}\Vert \varrho -\sigma\Vert_{1}, \label{bound-p-p-1}\\
\Vert \partial_{x}P_{\varrho} - \partial_{x}P_{\sigma}\Vert_{\infty} & \leqslant 8 \left(\frac{1}{ \lambda^{2}_{\min}(\widetilde{\varrho})} \Vert \partial_{x}\varrho - \partial_{x}\sigma\Vert_{1} +  2\frac{\big(\lambda_{\min}(\widetilde{\varrho}) + \lambda_{\min}(\widetilde{\sigma})\big)\Vert \partial_{x}\sigma\Vert_{1} }{\lambda^{2}_{\min}(\widetilde{\varrho}) \, \lambda^{2}_{\min}(\widetilde{\sigma})} \Vert \varrho -\sigma\Vert_{1}\right). \label{bound-dp-dp-1}
\end{align} 
\end{lemma}
\textit{Proof}. We remind that the eigenprojection $P_{\varrho}$ of a density matrix $\varrho$ on its support is given by \cite{book:operator,book:Hassani}
\begin{equation}
P_{\varrho} = -\frac{1}{2\pi i}\oint_{\Gamma} \mathrm{d}\lambda\,R_{\lambda}(\varrho),
\label{def:p-int}
\end{equation}
where $R_{\lambda}(\varrho)=(\varrho-\lambda\mathbbmss{I})^{-1}$ is the resolvent of $\varrho$, and $\Gamma$ is the contour of integration in the complex $\lambda$-plane ($\lambda\in\mathbbmss{C}$) which includes nonvanishing part of the spectrum of $\varrho$---that is, $\Sigma (\varrho)\backslash \{0\}\in \mathrm{interior}(\Gamma)$. Since $\Sigma (\varrho)\subseteq[0,1]$, we can choose $\Gamma$ to be a narrow strip as depicted in Fig. \ref{fig:boundary}. 

\begin{figure}[tp]
\includegraphics[scale=.4]{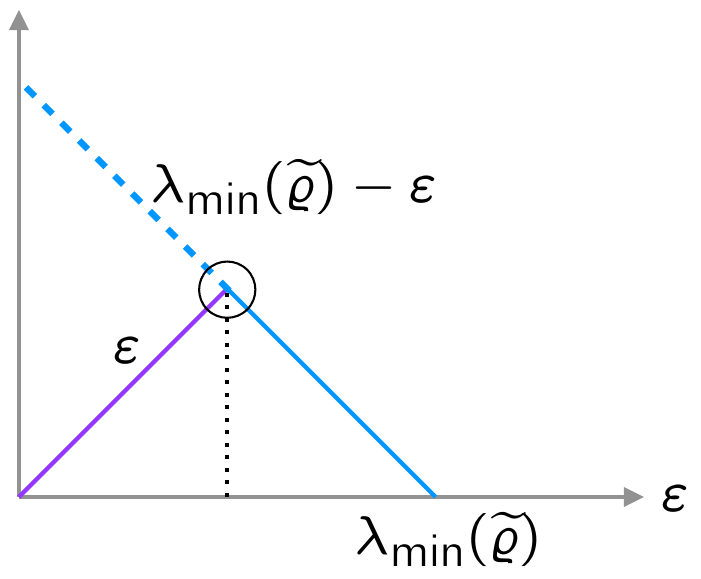}
\caption{Graphical proof that $\max_{\varepsilon\in [0,\lambda_{\min}(\widetilde{\varrho})]} \min\{\varepsilon,\lambda_{\min}(\widetilde{\varrho}) - \varepsilon\}=\lambda_{\min}(\widetilde{\varrho})/2$.}
\label{fig:min}
\end{figure}

Note that 
\begin{align}
P_{\varrho} - P_{\sigma} =&\frac{1}{2\pi i}\oint_{\Gamma}\mathrm{d}\lambda\big( R_{\lambda}(\sigma) - R_{\lambda}(\varrho) \big) \nonumber\\
= & \frac{1}{2\pi i}\oint_{\Gamma}\mathrm{d}\lambda\,R_{\lambda}(\sigma)\,(\sigma -\varrho)\,R_{\lambda}(\varrho),
\label{p1-p2}
\end{align}
where in the last line we used ``the second resolvent identity'' \cite{book:operator}
\begin{equation}
R_{\lambda}(\sigma) - R_{\lambda}(\varrho) = R_{\lambda}(\sigma)\,(\varrho - \sigma )\,R_{\lambda}(\varrho).
\label{2nd-res-id}
\end{equation}
Note that Eq. (\ref{p1-p2}) clearly indicates that when $\Vert \varrho - \sigma\Vert_{1}\to 0$ then $\Vert P_{\varrho} - P_{\sigma}\Vert_{\infty}\to0$. Now if $|\partial \Gamma|$ denotes the length of the contour $\Gamma$ and we employ the identity \cite{book:operator}
\begin{equation}
\Vert R_{\lambda}(X)\Vert_{\infty} =1/\mathrm{dist}(\lambda,\Sigma (X)),
\label{res-norm}
\end{equation}
we obtain
\begin{align}
\Vert P_{\varrho}-P_{\sigma}\Vert_{\infty} &\leqslant \frac{|\partial \Gamma|}{2\pi} \Vert \varrho-\sigma\Vert_{1} \,\max_{\lambda \in\Gamma}\frac{1}{\mathrm{dist}(\lambda,\Sigma (\varrho))} \, \max_{\lambda \in\Gamma}\frac{1}{\mathrm{dist}(\lambda,\Sigma (\sigma))}\nonumber\\
& = \frac{(1+\pi\varepsilon)}{\pi} \frac{\Vert \varrho - \sigma\Vert_{1}}{\min\{\varepsilon,\lambda_{\min}(\widetilde{\varrho}) - \varepsilon\}\,\min\{\varepsilon,\lambda_{\min}(\widetilde{\sigma}) - \varepsilon\}}   \nonumber\\
& \leqslant 2 \Vert \varrho -\sigma\Vert_{1} \frac{1}{\max_{\varepsilon\in [0,\lambda_{\min}(\widetilde{\varrho})]} \min\{\varepsilon,\lambda_{\min}(\widetilde{\varrho}) - \varepsilon\}}\,\frac{1}{\max_{\varepsilon\in [0,\lambda_{\min}(\widetilde{\sigma})]} \min\{\varepsilon,\lambda_{\min}(\widetilde{\sigma}) - \varepsilon\}} \nonumber\\
& \overset{\mathrm{Fig.\,\ref{fig:min}}}{=}\frac{8\Vert \varrho -\sigma\Vert_{1}}{\lambda_{\min}(\widetilde{\varrho}) \, \lambda_{\min}(\widetilde{\sigma})}.
\end{align}
Thus one can conclude that 
\begin{equation}
\Vert P_{\varrho}-P_{\sigma} \Vert_{\infty} \leqslant \min\left\{2,\frac{8}{\lambda_{\min}(\widetilde{\varrho}) \, \lambda_{\min}(\widetilde{\sigma})}\Vert \varrho -\sigma\Vert_{1} \right\}.
\label{bound-p-p-}
\end{equation}

From Eq. (\ref{def:p-int}) one can see that
\begin{equation}
\partial_{x}P_{\varrho} = \frac{1}{2\pi i}\oint_{\Gamma}\mathrm{d}\lambda\, R_{\lambda}(\varrho)\, \partial_{x}\varrho\, R_{\lambda}(\varrho),
\end{equation}
where we have used the identity $\partial_{x}(A^{-1})=-A^{-1} \partial_{x}A A^{-1}$ (for an $x$-dependent invertible and differentiable operator $A$), and have assumed that $\Gamma$ is an integral contour (akin to Fig. \ref{fig:boundary}) which encloses all nonvanishing eigenvalues of $\varrho(x)$ and $\varrho( x +\delta x )$ for sufficiently small (but nonvanishing) variations $\delta  x $. From this relation we conclude
\begin{align}
\Vert \partial_{x}P_{\varrho} \Vert_{\infty} & \leqslant \frac{|\partial\Gamma|}{2\pi} \max_{\lambda\in\Gamma} \big\Vert  R_{\lambda}(\varrho)\, \partial_{x}\varrho\, R_{\lambda}(\varrho)\big\Vert_{\infty}\nonumber\\
&\overset{(\ref{norm-1-1-inf})}{\leqslant} \frac{|\partial\Gamma|}{2\pi} \Vert \partial_{x}\varrho\Vert_{1} \,\max_{\lambda\in\Gamma}\Vert R_{\lambda}(\varrho)\Vert_{\infty}^{2} \nonumber\\
&\overset{\mathrm{(\ref{res-norm})}}{\leqslant} \frac{8\Vert \partial_{x}\varrho\Vert_{1}}{\lambda^{2}_{\min}(\widetilde{\varrho})}
\label{app:dP}
\end{align}
and
\begin{align}\label{bound-p-p-2}
\Vert \partial_{x}P_{\varrho} - \partial_{x}P_{\sigma}\Vert_{\infty} & \leqslant \frac{|\partial\Gamma|}{2\pi} \max_{\lambda\in\Gamma} \big\Vert  R_{\lambda}(\varrho)\, \partial_{x}\varrho\, R_{\lambda}(\varrho) - R_{\lambda}(\sigma)\, \partial_{x}\sigma\, R_{\lambda}(\sigma) \big\Vert_{\infty}\nonumber\\
&\overset{\mathrm{(\ref{important-ineq})}}{\leqslant} \frac{|\partial\Gamma|}{2\pi} \max_{\lambda\in\Gamma} \Big(\Vert \partial_{x}\varrho - \partial_{x}\sigma\Vert_{1}\, \Vert R_{\lambda}(\varrho)\Vert_{\infty}^{2} + \Vert R_{\lambda}(\varrho) - R_{\lambda}(\sigma)\Vert_{\infty} \Vert \partial_{x}\sigma\Vert_{1} (\Vert R_{\lambda}(\varrho)\Vert_{\infty} + \Vert R_{\lambda}(\sigma)\Vert_{\infty}) \Big)\nonumber\\
& \overset{\mathrm{(\ref{2nd-res-id}),\,(\ref{res-norm})}}{\leqslant} \frac{|\partial\Gamma|}{2\pi} \max_{\lambda\in\Gamma} \left( \frac{\Vert \partial_{x}\varrho - \partial_{x}\sigma\Vert_{1}}{\mathrm{dist}(\lambda,\Sigma (\varrho))^{2}} +  \Big[ \frac{1}{\mathrm{dist}(\lambda,\Sigma (\varrho))} + \frac{1}{\mathrm{dist}(\lambda,\Sigma (\sigma))} \Big] \frac{\Vert \partial_{x}\sigma\Vert_{1} \Vert \varrho -\sigma\Vert_{1}}{ \mathrm{dist}(\lambda,\Sigma (\varrho)) \, \mathrm{dist}(\lambda,\Sigma (\sigma))} \right) \nonumber\\
& \leqslant 8 \left(\frac{1}{ \lambda^{2}_{\min}(\widetilde{\varrho})} \Vert \partial_{x}\varrho - \partial_{x}\sigma\Vert_{1} +  2\frac{\big(\lambda_{\min}(\widetilde{\varrho}) + \lambda_{\min}(\widetilde{\sigma})\big)\Vert \partial_{x}\sigma\Vert_{1} }{\lambda^{2}_{\min}(\widetilde{\varrho}) \, \lambda^{2}_{\min}(\widetilde{\sigma})} \Vert \varrho -\sigma\Vert_{1}\right).
\end{align}
\hfill$\square$

\section{Bounds on the minimum eigenvalues of sum and product of two Hermitian operators}
\label{app:eig}
\subsection{$\lambda_{\min}(A+B)$}
\label{app:weyl}

Consider $ A $ and $ B $ to be two Hermitian $ n\times n $ matrices with eigenvalues $\lambda_{j}(A)$ and $\lambda_{j}(B)$. Assume that $ \lambda_{k}^{\downarrow}(A) $ denotes the eigenvalues of $A$ ordered decreasingly, that is, $ \lambda_{1}^{\downarrow}(A)\equiv \lambda_{\max}(A) \geqslant \ldots\geqslant \lambda_{n}^{\downarrow}(A) \equiv \lambda_{\min}(A)$. The Weyl inequality yields \cite{book:Bhatia}
\begin{align}
\lambda_{j}^{\downarrow}(A+B)\leqslant\lambda_{i}^{\downarrow}(A)+\lambda_{j-i+1}^{\downarrow}(B),~~~~~~~~~i\leqslant j,\\
\lambda_{j}^{\downarrow}(A+B)\geqslant\lambda_{i}^{\downarrow}(A)+\lambda_{j-i+n}^{\downarrow}(B),~~~~~~~~~i\geqslant j.
\end{align}
Choosing $ i=j=n $ gives
\begin{equation}
\lambda_{\min}(A+B)\geqslant\lambda_{\min}(A)+\lambda_{\min}(B).
\label{wy}
\end{equation}

\subsection{$\lambda_{\min}(AB)$}
\label{ap3}

Let us assume $\mathbf{x}=(x_{1},\ldots,x_{n})$, $\mathbf{y}=(y_{1},\ldots,y_{n})$, and $\mathbf{z}=(z_{1},\ldots,z_{n})$ are $n$-vectors with nonnegative elements. From majorization theory \cite{book:Bhatia} we recall the definitions
\begin{align}
&\log \mathbf{x} \prec_{w}\log \mathbf{x} ~~~~~\mathrm{if}~~~~~\prod_{i=1}^{k}{x^{\downarrow}_{i}}\leqslant\prod_{i=1}^{k}{y^{\downarrow}_{i}}; ~~~~~\mathrm{for}~~~k=1,2,\ldots , n\\
&\log \mathbf{x} \prec\log \mathbf{y} ~~~~~\mathrm{if}~~~~~\log \mathbf{x} \prec_{w}\log \mathbf{y}~~~~~\mathrm{with}~~~~~\prod_{i=1}^{n}{x^{\downarrow}_{i}}=\prod_{i=1}^{n}{y^{\downarrow}_{i}}.
\end{align}

Now we use Corollary 3.4.6 (Lidskii) in Ref. \cite{book:Bhatia}. Let $ A $ and $ B $ be two positive operators. Then all eigenvalues of $ AB $ are nonnegative and 
\begin{equation}
\log\bm{\lambda}^{\downarrow}(A)+\log\bm{\lambda}^{\uparrow}(B)\prec\log\bm{\lambda}(AB)\prec\log\bm{\lambda}^{\downarrow}(A)+\log\bm{\lambda}^{\downarrow}(B),
\end{equation}
where $\bm{\lambda}^{\downarrow}(A)$ ($\bm{\lambda}^{\uparrow}(A)$) denotes the vector of the eigenvalues of $A$ ordered decreasingly (increasingly). Thus
\begin{align}
&\prod_{i=1}^{k}{\lambda^{\downarrow}_{i}(AB)}\leqslant\prod_{i=1}^{k}{\lambda^{\downarrow}_{i}(A)}\prod_{i=1}^{k}{\lambda^{\downarrow}_{i}(B)}\label{start2},\\
&\prod_{i=1}^{n}{\lambda^{\downarrow}_{i}(AB)}=\prod_{i=1}^{n}{\lambda^{\downarrow}_{i}(A)}\prod_{i=1}^{n}{\lambda^{\downarrow}_{i}(B)}.
\label{start}
\end{align}
We can write Eq. \eqref{start} as
\begin{align}
\lambda_{\min}(AB)\prod_{i=1}^{n-1}{\lambda^{\downarrow}_{i}(AB)}&=\lambda_{\min}(A)\prod_{i=1}^{n-1}{\lambda^{\downarrow}_{i}(A)}\, \lambda_{\min}(B)\prod_{i=1}^{n-1}{\lambda^{\downarrow}_{i}(B)}\nonumber\\
\frac{\lambda_{\min}(AB)}{\lambda_{\min}(A)\, \lambda_{\min}(B)}&=\frac{\prod_{i=1}^{n-1}{\lambda^{\downarrow}_{i}(A)}\prod_{i=1}^{n-1}{\lambda^{\downarrow}_{i}(B)}}{\prod_{i=1}^{n-1}{\lambda^{\downarrow}_{i}(AB)}} \overset{\mathrm{(\ref{start2})}}{\geqslant} 1.
\end{align}
Hence, we obtain 
\begin{equation}
\lambda_{\min}(AB)\geqslant\lambda_{\min}(A)\,\lambda_{\min}(B).
\label{min-AB}
\end{equation}

\subsection{$\lambda_{\min}(\mathpzc{E}[\varrho_{0}])$}
\label{}

Assume a completely-positive trace preserving dynamical map $\mathpzc{E}$ applied on an initial state $\varrho_{0}$, $\mathpzc{E}[\varrho_0]=\sum_k A_k \varrho_0 A_k^{\dag}$. We have 
\begin{align}
\lambda_{\min}\big(\textstyle{\sum_{k}} A_{k}\varrho_{0}A^{\dag}_{k}\big) & \overset{\mathrm{(\ref{wy})}}{\geqslant} \sum_{k}\lambda_{\min}\big(A_{k}\varrho_{0} A_{k}^{\dag}\big),\nonumber\\
&=\sum_{k} \lambda_{\min}\big(\varrho_{0} A_{k}^{\dag} A_{k}\big), \nonumber\\
&\overset{\mathrm{(\ref{min-AB})}}{\geqslant} \lambda_{\min}(\varrho_{0})\sum_{k} \lambda_{\min}\big(A_{k}^{\dag} A_{k}\big),
\end{align}
where in the second line we have used the identity $\lambda(AB)=\lambda(BA)$. Positivity of $A^{\dag}_{k} A_{k}$ and the trace preserving condition $\sum_{k}A^{\dag}_{k} A_{k}=\mathbbmss{I}$ imply that $0\leqslant \lambda\big(A_{k}^{\dag} A_{k}\big) \leqslant 1$.

For random unitary channels of the form $\mathpzc{U}[\circ]=\sum_{k} p_{k} U_{k} \circ U^{\dag}_{k}$, where $\{p_{k}\}$ constitutes a probability distribution and $U_{k}$s are unitary, the above bound gives
\begin{equation}
\lambda_{\min}\big(\textstyle{\sum_{k}} p_{k} U_{k}\varrho_{0}U^{\dag}_{k}\big) \geqslant \lambda_{\min}(\varrho_{0}).
\label{bnd}
\end{equation}

As an example, for a $d$-dimensional depolarizing channel $\mathpzc{U}[\varrho_{0}]=(1-p)\varrho_{0} + (p/d)\mathbbmss{I}$, we have $\lambda_{\min}(\mathpzc{U}[\varrho_{0}])=(1-p)\lambda_{\min}(\varrho_{0}) +p/d$, which is obviously $\geqslant \lambda_{\min}(\varrho_{0})$---in agreement with Eq. (\ref{bnd}).

\section{Proof of Eq. (\ref{mosh-exp-dif})}
\label{app:eq-proof}

Note that
\begin{align}
\Vert \mosh\varrho - \mosh\sigma\Vert_{1}&=\Big\Vert\int_{0}^{1}\mathrm{d}t~e^{tH}~\mosh H~e^{(1-t)H}-e^{tG }~\mosh G ~e^{(1-t)G }\Big\Vert_{1}\nonumber\\
&\leqslant\int_{0}^{1}\mathrm{d}t~\Vert \underbrace{e^{tH}}_{A}~\underbrace{\mosh H~e^{(1-t)H}}_{B}-\underbrace{e^{tG }}_{A^{\prime}}~\underbrace{\mosh G ~e^{(1-t)G }}_{B^{\prime}}\Vert_{1}\nonumber\\
&\overset{\mathrm{(\ref{imp-ineq})}}{\leqslant}\int_{0}^{1}\mathrm{d}t\, \Big(\Vert e^{tH}-e^{tG }\Vert_{\infty} \, \Vert\mosh H~e^{(1-t)H}\Vert_{1} +\Vert e^{tG }\Vert_{\infty}\, \Vert \mosh He^{(1-t)H}-\mosh G e^{(1-t)G }\Vert_{1}\Big)\nonumber\\
&=\int_{0}^{1}\mathrm{d}t~\Vert e^{tH}-e^{tG }\Vert_{\infty}\, \Vert\mosh H~e^{(1-t)H}\Vert_{1} +\int_{0}^{1}\mathrm{d}t~\Vert e^{tG }\Vert_{\infty} \,\Vert \mosh He^{(1-t)H}-\mosh G e^{(1-t)G }\Vert_{1} .
\label{mosh-exp-dif-1}
\end{align}
Let us consider the two terms in Eq. (\ref{mosh-exp-dif-1}) separately. For the first term we have
\begin{align}
\int_{0}^{1}\mathrm{d}t~\Vert e^{tH}-e^{tG }\Vert_{\infty} \,\Vert\mosh H~e^{(1-t)H}\Vert_{1} & \overset{(\ref{norm-1-1-inf})}{\leqslant} \int_{0}^{1}\mathrm{d}t~\Vert e^{tH}-e^{tG }\Vert_{\infty}\,\Vert\mosh H\Vert_{1}~\Vert e^{(1-t)H}\Vert\nonumber\\
&\overset{(\ref{aa})}{\leqslant}\int_{0}^{1}\mathrm{d}t~\Vert t(H-G )\Vert_{\infty} \int_{0}^{1} \mathrm{d}\tau\,\Vert e^{\tau tH}\Vert_{\infty}\, \Vert e^{(1-\tau)tG }\Vert_{\infty}\, \Vert\mosh H\Vert_{1}\,\Vert e^{(1-t)H}\Vert_{\infty}\nonumber\\
&=\int_{0}^{1}\mathrm{d}t~t\Vert H-G \Vert_{\infty} \int_{0}^{1}\mathrm{d}\tau\,\Vert \varrho^{\tau t}\Vert_{\infty} \,\Vert\sigma ^{(1-\tau)t}\Vert_{\infty}\, \Vert\mosh H\Vert_{1}\, \Vert \varrho ^{1-t}\Vert_{\infty} \nonumber\\
&\overset{\Vert \varrho^{s}\Vert \leqslant 1\,\mathrm{for}\,0\leqslant s\leqslant 1}{\leqslant} \frac{1}{2}\Vert H-G \Vert_{\infty} \, \Vert\mosh H\Vert_{1} .
\label{1-mosh-dif}
\end{align}
Similarly, for the second term in Eq. (\ref{mosh-exp-dif-1}) one can obtain
\begin{align}
&~~~\int_{0}^{1}\mathrm{d}t~\Vert e^{tG }\Vert_{\infty}\, \Vert \underbrace{\mosh H}_{A}\underbrace{e^{(1-t)H}}_{B}-\underbrace{\mosh G }_{A^{\prime}}\underbrace{e^{(1-t)G }}_{B^{\prime}}\Vert_{1}\nonumber\\
&\overset{(\ref{imp-ineq})}{\leqslant}\int_{0}^{1}\mathrm{d}t~\Vert e^{tG }\Vert_{\infty} \left( \Vert\mosh H-\mosh G \Vert_{1}\,\Vert e^{(1-t)H}\Vert_{\infty} +\Vert\mosh G \Vert_{1}\, \Vert e^{(1-t)H}- e^{(1-t)G }\Vert_{\infty} \right) \nonumber\\
&\overset{(\ref{aa})}{\leqslant}\int_{0}^{1}\mathrm{d}t~\Vert e^{tG }\Vert_{\infty} \left( \Vert\mosh H-\mosh G \Vert_{1} \,\Vert e^{(1-t)H}\Vert_{\infty} +(1-t)\Vert\mosh G \Vert_{1}\, \Vert H-G\Vert_{\infty} \int_{0}^{1}\mathrm{d}\tau\,\Vert e^{\tau(1-t)H}\Vert_{\infty}\, \Vert e^{(1-\tau)(1-t)G }\Vert_{\infty}\right) \nonumber\\
&=\int_{0}^{1}\mathrm{d}t~\Vert \sigma ^{t}\Vert_{\infty} \left( \Vert\mosh H-\mosh G \Vert_{1}\, \Vert \varrho ^{1-t}\Vert_{\infty} + (1-t)\Vert\mosh G \Vert_{1}\, \Vert H-G \Vert_{\infty} \int_{0}^{1}\mathrm{d}\tau\, \Vert \varrho ^{\tau(1-t)}\Vert_{\infty} \,\Vert \sigma ^{(1-\tau)(1-t)}\Vert_{\infty}\right) \nonumber\\
&\leqslant\Vert\mosh H-\mosh G \Vert_{1} +\frac{1}{2}\Vert\mosh G \Vert_{1} \,\Vert H-G \Vert_{\infty}.
\label{2-mosh-dif}
\end{align}
Substituting Eqs. (\ref{1-mosh-dif}) and (\ref{2-mosh-dif}) in Eq. (\ref{mosh-exp-dif-1}) yields
\begin{align}
\Vert \mosh\varrho - \mosh\sigma\Vert_{1}\leqslant\frac{1}{2}\Vert H-G \Vert_{\infty} \big(\Vert\mosh H\Vert_{1} +\Vert\mosh G \Vert _{1}\big) +\Vert\mosh H-\mosh G \Vert_{1} .
\end{align}
\hfill$\square$


\end{document}